\documentclass[pdflatex,sn-mathphys-num]{sn-jnl}
\usepackage{graphicx}%
\usepackage{multirow}%
\usepackage{amsmath,amssymb,amsfonts}%
\usepackage{amsthm}%
\usepackage{mathrsfs}%
\usepackage[title]{appendix}%
\usepackage{xcolor}%
\usepackage{textcomp}%
\usepackage{manyfoot}%
\usepackage{booktabs}%
\usepackage{algorithm}%
\usepackage{algorithmicx}%
\usepackage{algpseudocode}%
\usepackage{listings}%
\usepackage{siunitx}%
\usepackage{hyperref}%
\usepackage{lmodern}%
\unnumbered
\begin{document}

\title[Article Title]{QARPET: A Crossbar Chip for Benchmarking Semiconductor Spin Qubits}

\author[1]{\fnm{Alberto} \sur{Tosato}}
\author[1]{\fnm{Asser} \sur{Elsayed}}
\author[1]{\fnm{Federico} \sur{Poggiali}}
\author[1]{\fnm{Lucas} \sur{Stehouwer}}
\author[1]{\fnm{Davide} \sur{Costa}}
\author[1]{\fnm{Karina} \sur{Hudson}}
\author[1]{\fnm{Davide} \sur{Degli Esposti}}
\author*[1]{\fnm{Giordano} \sur{Scappucci}}\email{g.scappucci@tudelft.nl}

\affil[1]{\orgdiv{QuTech and Kavli Institute of Nanoscience}, \orgname{Delft University of Technology}, \orgaddress{\street{Lorentzweg 1}, \postcode{2628 CJ} \city{Delft}, \country{The Netherlands}}}

\abstract{Large-scale integration of semiconductor spin qubits into quantum processors hinges on the ability to characterize quantum components at scale, a task challenged by their operation at sub-kelvin temperatures, in the presence of magnetic fields, and by the use of radio-frequency signals. Here, we present QARPET (Qubit-Array Research Platform for Engineering and Testing), a scalable architecture for characterizing spin qubits using a quantum dot crossbar array. The crossbar features tightly-pitched spin qubit tiles and is implemented in planar germanium, by fabricating a large device with the potential to host 1058 hole spin qubits. Measurements on a patch of 40 tiles demonstrate key device functionality at millikelvin temperatures, including tile addressability, threshold voltage and charge noise statistics, and the characterisation of hole spin qubits and their coherence times in a single tile. These demonstrations pave the way for a new generation of quantum devices designed for the statistical characterisation of spin qubits.}

\maketitle

\section{Introduction}\label{sec1}

The recent demonstration of spin qubits~\cite{burkard2023semiconductor} manufactured in a modern semiconductor foundry~\cite{zwerver2022qubits} offers a promising avenue to address scalability challenges of quantum technology, by leveraging decades of technology development in the semiconductor industry. However, integrating millions of highly coherent spin qubits into a quantum processor still demands substantial industrial developments to advance material synthesis, fabrication processes, and control strategies~\cite{de_leon_materials_2021,saraiva_materials_2022}. These developments depend critically on the ability to test quantum components at scale for yield and performance, under environmental conditions, such as cryogenic temperatures, that differ significantly from those in which current semiconductor technology operates. 

Various approaches have been explored to streamline the cryogenic testing of quantum devices, including on-chip and off-chip multiplexers~\cite{ward_integration_2013,al-taie_cryogenic_2013,puddy_multiplexed_2015,schaal_cmos_2019,paquelet_wuetz_multiplexed_2020,pauka_characterizing_2020,ruffino_cryo-cmos_2021,bavdaz_quantum_2022,wolfe_-chip_2024,thomas_rapid_2025} to improve the limited input/output connectors in existing cryostats. Alternatively, a cryogenic \SI{300}{\milli\meter} wafer prober has been employed to perform low-frequency measurements on a large number of quantum dots at \SI{1.6}{\kelvin}~\cite{Neyens2024ProbingWafers}, offering fast feedback to optimise CMOS-compatible fabrication processes of spin qubit devices.
However, none of these approaches currently provides a scalable solution for statistical measurements of spin qubits, which typically require radio-frequency (RF) measurements at millikelvin temperatures in the presence of magnetic fields. 

In this article, we introduce a scalable architecture for characterising spin qubits using a quantum dot crossbar array, named QARPET (Qubit-Array Research Platform for Engineering and Testing). The crossbar is based on arrayed, individually addressable, and tightly pitched spin qubit tiles, featuring a qubit density of \SI{2E6}{\per\square\milli\meter} and sublinear scaling of interconnects. This architecture draws inspiration from device matrix arrays (DMAs) widely used in the semiconductor industry as test vehicles for assessing the matching properties of transistors\cite{pelgrom_matching_1989,ohkawa_analysis_2003,tsunomura_analysis_2009,mizutani_measuring_2011}. We implement QARPET in planar germanium quantum wells~\cite{scappucci_germanium_2021} and fabricate a large crossbar array device of $23\times23$ tiles, which offers the potential to test 1058 single hole spin qubits within a single cool-down. In this first implementation, we demonstrate the unique tile addressability and the capability to acquire spin qubit device metrics, such as threshold voltages and charge noise, by using RF reflectometry at millikelvin temperatures. These measurements are extended to a statistical analysis across 40 tiles, showcasing the scalability of the approach. As a proof of principle, we demonstrate spin control within a tile by implementing singlet-triplet qubits, operated with baseband-only control signals, and Loss-DiVincenzo single-hole qubits, driven by electric dipole spin resonance (EDSR) and characterize their coherence time.

\newpage

\section{Results}\label{sec2}
\subsection{A scalable spin qubit tile in a crossbar array}\label{subsec1}
\begin{figure}
\centering
\includegraphics[width=1\textwidth]{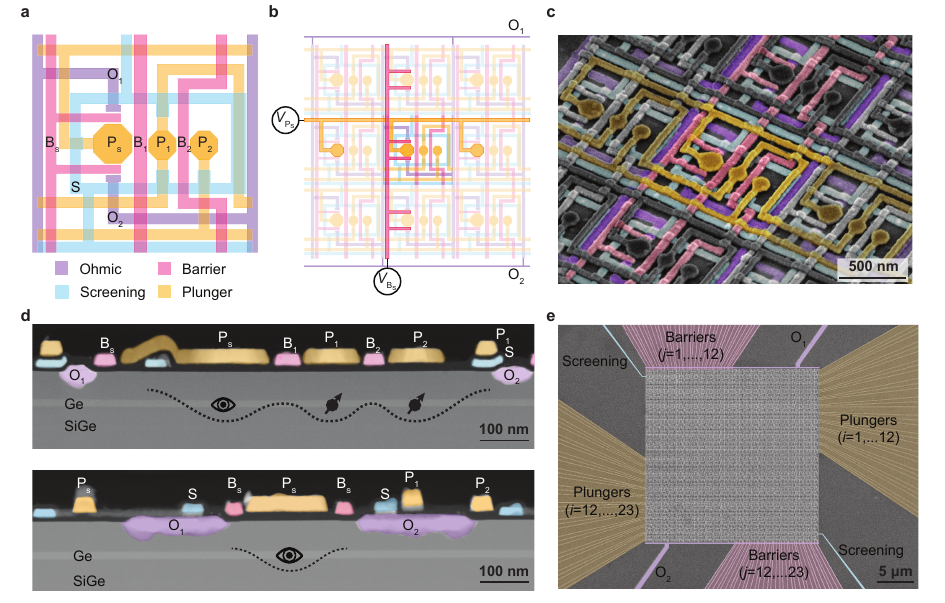}
\caption{ \textbf{Quantum dot spin-qubit crossbar}.
        \textbf{a,} Design of a scalable spin-qubit tile comprising a sensing dot and a double dot, with plunger gates ($P_\mathrm{s}$, $P_1$, $P_2$), barrier gates ($B_\mathrm{s}$, $B_1$, $B_2$), ohmic contacts ($O_1$, $O_2$), and a screening gate {$S$}.
        \textbf{b,} The tile layout, with meandering vertical plunger gates and horizontal barrier gates, allows for integration into a scalable device architecture. A single tile, e.g. the central tile in the illustrated 3$\times$3 array, is selected for measurements by energizing the correspondent sensor plunger and barrier with gate bias $V_{P_\mathrm{s}}$ and $V_{B_\mathrm{s}}$ respectively. The reflectance measured between $O_1$ and $O_2$ will be proportional to the reflectance of the sensor in the central tile, allowing to be exclusively sensitive to the charges in the dots of the selected tile. 
        \textbf{c,} False coloured scanning electron microscopy image highlighting a tile within a crossbar device comprising 23$\times$23 tiles and fabricated on a Ge/SiGe heterostructure following the architecture in \textbf{a},\textbf{b}. The colour scheme matches the schematics in panel~\textbf{a}. The tile has a footprint of $1\times1$~\unit{\square\micro\meter}, the circular plungers defining the sensor and the dots have a nominal diameter of \SI{180}{nm} and \SI{130}{nm}, respectively. The barriers $B_1$ and $B_2$ have a width of \SI{60}{nm} and \SI{50}{nm}, respectively. 
        \textbf{d,} Transmission electron microscope images showing cross-sections of a tile within the crossbar. The top panel is along the horizontal axis of the tile, crossing the circular sensor and dot plungers. The bottom panel is along the vertical axis of the tile, crossing the sensor barriers and circular plunger. The multi-layer gate structure is fabricated on a buried \SI{16}{nm} thick Ge/SiGe quantum well, positioned at \SI{55}{nm} from the surface. The colour scheme and labels match the schematics in ~\textbf{a}.
        \textbf{e,} Scanning electron micrograph image showing the entire crossbar device extending over an area of $23\times23$~\unit{\square\micro\meter} and featuring 529 tiles arranged in a 23$\times$23 array indexed by plunger rows $i$ and barrier columns $j$. The gate electrodes and the ohmic contacts fan out at the periphery of the crossbar.
        }
\label{fig1}
\end{figure}

Figure \ref{fig1}a,b illustrates the design of a scalable spin qubit tile that can be arranged into an $n\times m$ crossbar array architecture. Plunger gates control the chemical potentials of the charge sensor ($P_\mathrm{s}$) and of two quantum dots ($P_1$, $P_2$). Barrier gates adjust the coupling between the sensor and the ohmics ($B_\mathrm{s}$), the sensor and the neighbouring dot ($B_1$), and between the dots ($B_2$). A global screening gate ($S$) shapes the surrounding potential landscape. Two ohmic electrodes ($O_1$, $O_2$) run vertically and merge at the edges into a single pair of ohmic contacts for the entire crossbar, minimizing the off-chip resonators needed for RF-reflectometry. Rows of tiles share plunger gates, columns share barrier gates, and the meandering design of $P_1$ prevents overlaps, enabling $P_\mathrm{s}$, $P_1$, and $P_2$ to be patterned in the same layer to simplify fabrication. Similarly to wordlines and bitlines in random-access memories~\cite{veldhorst_silicon_2017,li_crossbar_2018}, a specific tile indexed $(i,j)$ is activated and addressed by energising its $P_\mathrm{s}$ and $B_\mathrm{s}$ electrodes, which form a sensor dot sufficiently coupled to the ohmic contacts to provide a transport path. With sensor-based tile selectivity, dot plungers and barriers electrodes can be shorted across tiles, reducing the required control lines. An $n \times m$ crossbar hosting $2mn$ qubits needs only $(n+m+7)$ lines: $(n+m)$ for $B_\mathrm{s}$ and $P_\mathrm{s}$, seven for $P_1$, $P_2$, $B_1$, $B_2$, $S$, $O_1$, and $O_2$. Therefore, this architecture achieves sub-linear scaling of control lines with number of qubits, following Rent’s rule with an exponent $p=0.5$~\cite{Landman1971OnGraphs, Franke2019RentsComputing}.

We implement this architecture in a low-disorder Ge/SiGe heterostructure on a Si wafer~\cite{lodari_low_2021} (Methods), which was used in several spin qubit experiments~\cite{hendrickx_single-hole_2020,hendrickx2024sweet,Wang2024OperatingSpins,zhang_universal_2024}. We fabricate a crossbar array of $23\times23$ spin qubit tiles, which can support up to 1058 individually addressable spin qubits, while requiring only 53 control lines. The fabrication process entails germanosilicide ohmic contacts to the germanium quantum well and a multi-layer gate stack patterned by electron-beam lithography and metal lift-off (Methods). The scanning electron microscope image in Fig.~\ref{fig1}c provides detailed views of a tile within the crossbar. The tile footprint is
\SI{1}{\micro\meter}, achieved through a tightly knit fabric of nanoscale electrodes and yielding a high density of quantum dot qubits of \SI{2E6}{\per\square\milli\meter}. Furthermore, the tile footprint is comparable to the length scale of strain and compositional fluctuations of the heterostructure~\cite{stehouwer2023germanium}, making the device suitable for probing variations in quantum dot metrics arising from the underlying heterostructure. 

The transmission electron microscope images in Fig.~\ref{fig1}d show cross sections of a tile along the circular dot and sensor plungers direction (top panel) and, orthogonally, across the sensor plunger and barrier direction (bottom panel). These images illustrate the germanosilicide ohmic contacts to the buried germanium quantum well and the three layers of gates (screening, barriers, plungers) with dielectric in between. To electrostatically define charge sensors and quantum dots in the buried Ge quantum well, circular plunger gates are set to negative potential to accumulate holes, while barrier gates potentials are adjusted to tune the tunnel couplings between source and drain reservoirs and between the dots.
The image of the entire crossbar in Fig~\ref{fig1}e highlights the fanout of the nanoscale gate electrodes and ohmic contacts at the periphery of the crossbar (see also Supplementary Fig.~1). The realization of such QARPET chip demonstrates the viability of our approach to array dense spin-qubit tiles even without the strict process control available in an advanced semiconductor foundry.

\subsection{Charge sensor addressability and single hole occupancy}\label{subsec2}

\begin{figure}
\centering
\includegraphics[width=.98\textwidth]{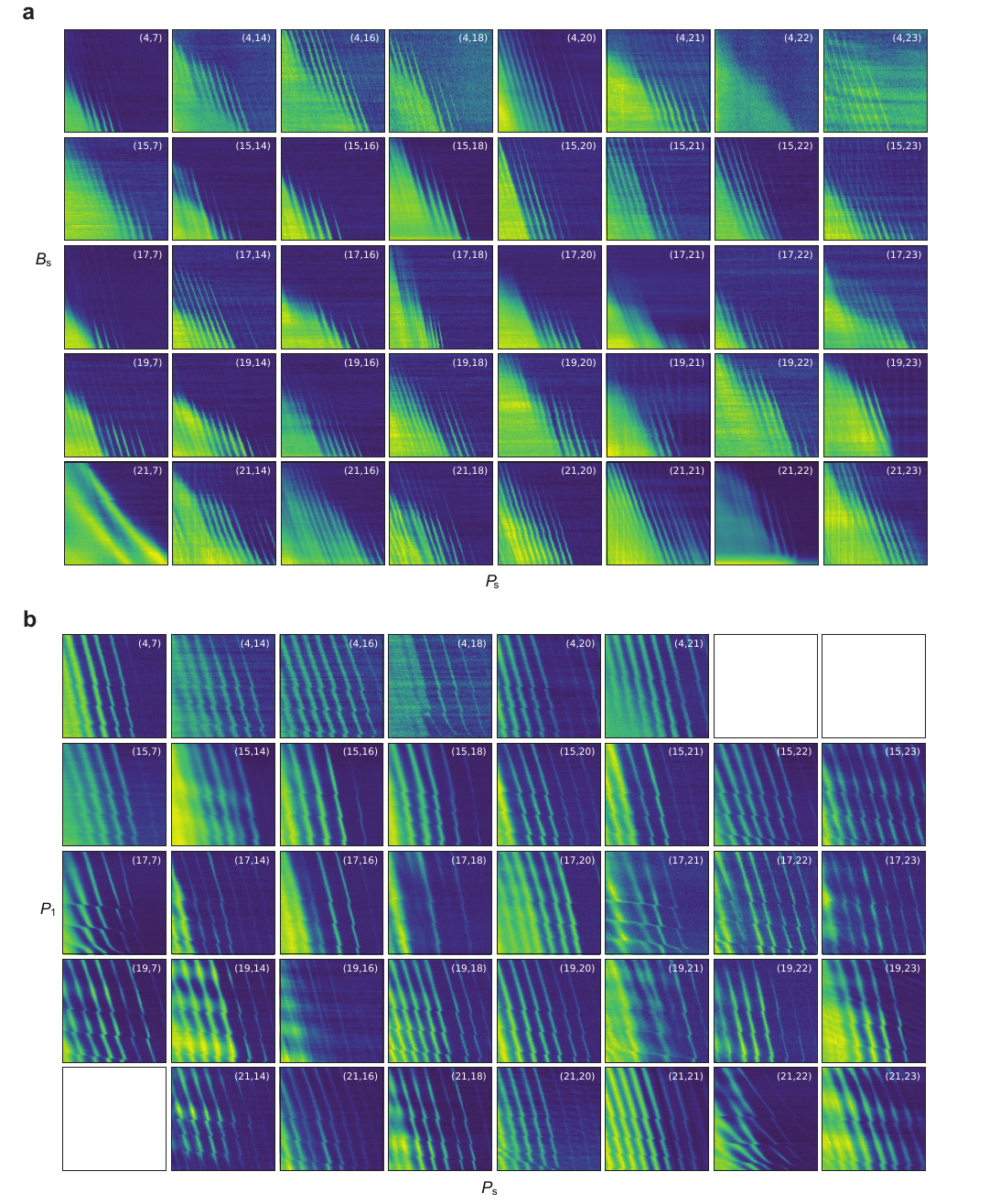}
\caption{ \textbf{Addressability and single hole quantum dot regime.} \textbf{a,} The sensor reflectance maps are measured on a QARPET device with RF reflectometry as a function of the sensor plunger ($P_\mathrm{s}$) and the sensor barrier ($B_\mathrm{s}$) for a subset of 40 tiles arranged in 5 rows and 8 columns of the device. The selectivity of each tile is confirmed by the observation that the sensor signal of all sensor-barrier maps presents Coulomb peaks with a negative slope. This indicates that the sensing dot is capacitively coupled to both the intended sensor plunger gate ($P_\mathrm{s}$) and the sensor barrier gate ($B_\mathrm{s}$). Furthermore, Coulomb blockade is observed in all tiles expect for two ((4,22) and ([21,7)), indicating that the tile design is suitable for performing charge sensing. \textbf{b,} Sensor reflectance maps as a function of the sensor plunger ($P_\mathrm{s}$) and the plunger of dot 1 ($P_1$) show charge occupation down to the last hole for the 37 tiles with a sensor that displays a clear Coulomb blockade signal. Supplementary Figs.~2,3 include color-scale bar and voltage values.}
\label{fig2}
\end{figure}

We evaluate the functionality of the device at \SI{100}{\milli\kelvin}, by measuring a subset of 40 tiles, arranged in five rows and eight columns. The pre-selection process, based on cryogenic setup and packaging constraints and testing at \SI{4.2}{\kelvin}, is detailed in Supplementary Fig.\,1, together with considerations of potential failure modes of the architecture, as discussed in Supplementary Note\,1. 
Tile selectivity is demonstrated using RF-reflectometry to measure sensor reflectance through the ohmic contacts, tuning the sensor of one tile at a time to a regime showing clear Coulomb blockade signatures in $B_\mathrm{s}$~versus~$P_\mathrm{s}$ gate maps (Fig.~\ref{fig2}a). 
A negative slope of the Coulomb peaks confirms that the measured reflectance corresponds to the targeted tile. If the reflectance signal originated from another tile, vertical or horizontal Coulomb peak lines would appear. Vertical lines would indicate the signal comes from the sensor of a tile in the same row as the targeted tile, as the selected barrier gate would not affect its chemical potential, while horizontal lines would correspond to a sensor of a tile in the same column. We are able to tune the sensor of 38/40 tiles in Coulomb blockade, demonstrating single tile addressability in the dense array. These measurements also demonstrate the robustness of the RF-reflectometry approach in addressing hundreds of sensor dots connected in parallel, despite the increased parasitic capacitance.

Using charge sensing, we demonstrate quantum dots in the few-hole occupation regime, a typical condition for spin qubit operations. We focus on the occupation of the first dot ($D_1$) that can be directly loaded from the nearby sensing dot easing the operation. Tuning of dot 2 ($D_2$) in this device was prevented by barrier $B_2$ leaking to ground.
We sequentially tune $D_1$ to the few-hole regime in each tile by measuring charge stability diagrams of $P_\mathrm{s}$~versus~$P_1$ (Fig.~\ref{fig2}b) and adjusting voltages in real time, with all other gates grounded during tuning.  Overall, we tune $D_1$  to the last hole in 37/40 tiles, with the first transition line approximately centred in each stability diagram. Only three dots failed achieving the last hole due to sensor issues, with tile (4,23) showing insufficient contrast in sensor Coulomb peaks and tiles (4,22) and (23,7) failing to reach Coulomb blockade within the gate voltage boundaries set for this measurement (see Supplementary Figs.\,2,3). This failure mode does not affect the operation of other tiles, other failure modes for this device type are discussed in Supplementary Note\,1.

\subsection{Electrostatic variability}\label{subsec3}

We investigate the electrostatic variability across the tiles by focusing on lever arms, single-hole voltages, and addition voltages related to $D_1$. The distributions of these metrics across different tiles reflect differences in dot shape and position, the uniformity of the semiconductor heterostructure and gate stack, and must be evaluated collectively to account for tile-specific tuning conditions.

\begin{figure}
\centering
\includegraphics[width=1\textwidth]{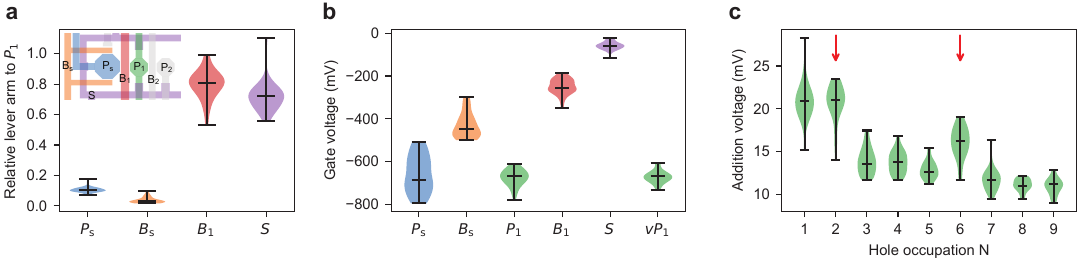}
    \caption{
        \textbf{Electrostatic variability.}
    \textbf{a,} By analysing statistical data over multiple tiles of a QARPET device such as in Fig.\ref{fig2} and Supplementary Figs.\,4--7, we obtain violin plots of the relative lever arm of all gates with respect to $P_1$. 
    \textbf{b,} The violin plots show  the distribution of voltages for the different gates whilst the last hole is reached in dot 1. We also report the values of virtual $P_1$ ($\text{v}P_1$), calculated by adding the contribution of each surrounding gate, weighted by the relative lever arm and normalized by the median value of $P_1$. While part of the variability of $P_1$ arises from the different voltages applied to the surrounding gates, $\text{v}P_1$ accounts for this and therefore results in a smaller variability. The variability in $\text{v}P_1$ can be used across different crossbar devices as a metric to benchmark, for example, improvements in the uniformity of the Ge/SiGe heterostructure.
    \textbf{c,} The violin plots show the distribution of the addition voltage for different hole occupations of dot 1. We observe on average a larger addition voltage for N=2 and 6 (red arrows), consistent with shell filling of circular hole quantum dots.
    }
\label{fig3}
\end{figure}

Figure~\ref{fig3}a shows the distributions of the lever arms of nearby gates ($P_s$, $B_s$, $S$, and $B_1$) relative to $D_1$, divided by the lever arm of $P_1$ to $D_1$ (Methods and Supplementary Figs.\,4--7)~\cite{Tidjani2023VerticalWell}.
As expected from the tile design (Fig.~\ref{fig3}a, inset), the chemical potential of $D_1$ is more influenced by the gates closest to the dot plunger ($B_1$ and $S$) and less by sensor gates ($P_s$ and $B_s$) in spite of their larger dimensions.
Figure~\ref{fig3}b shows the distribution of voltages applied to gates within a tile for tuning $D_1$ to the single-hole regime. The distribution for the sensor gates has the largest standard deviation ($\sigma_{P_s}=86$~mV, $\sigma_{B_s}=58$~mV) compared to the other gates (e.g. $\sigma_{P_1}=38$~mV, $\sigma_{B_1}=36$~mV). This is expected given the manual tuning approach and the typically broad voltage range available for achieving sharp Coulomb peaks that can be used for sensing.
To isolate the variability introduced by manual tuning, we calculate the virtual gate voltage $\mathrm{v}P_1$ (proportional to the chemical potential of $D_1$, Methods) by summing the contributions of each gate weighted by its relative lever arm. We observe that the distribution width is reduced to $\sigma_{vP_1} = 29$~mV compared to $\sigma_{P_1} = 38$~mV. Additionally, we follow the analysis in Refs.~\cite{Neyens2024ProbingWafers,ha_flexible_2022} and calculate the standard deviation of the plunger and barrier voltage difference $\sigma(P_1 - P_b)$ for achieving the first hole. We obtain $\sigma(P_1 - B_1) = 63~\mathrm{mV}$ pointing to a uniform disorder potential environment across the device (see Supplementary Fig.\,8).

In Fig.~\ref{fig3}c we investigate the distribution of addition voltages across different tiles for different hole occupations. 
We observe on average a larger addition voltage for occupation with two and six holes (red arrows), consistent with shell filling of circular hole quantum dots~\cite{VanRiggelen2021ADots} and absence of low-energy excited states in all the dots~\cite{lim_spin_2011}. From the distribution, we note that the variability of addition voltage for each occupation is about 10\% of its mean value, which gives an insight into the minimum expected variability of pulse amplitudes required for operating multiple qubits in a device.
Comparing the median addition voltage for the first hole ($\sim$21 mV) to the distribution of $P_1$ voltages for the first charge transition (Fig.~\ref{fig3}b), we estimate that applying the $P_1$ median voltage to different tiles would tune $D_1$ to single-hole occupation in 25\% of the cases. This increases to 57\% and 75\% when targeting occupation up to three and five holes, respectively. Odd charge occupation different than the single-hole can be explored for robust and localized qubit control~\cite{john_two-dimensional_2025}. 
These insights provides a concrete benchmark for further materials and process optimization towards shared-control spin-qubit architectures~\cite{borsoi_shared_2024}.

\subsection{Charge noise}\label{subsec4}
\begin{figure}
\centering
\includegraphics[width=1\textwidth]{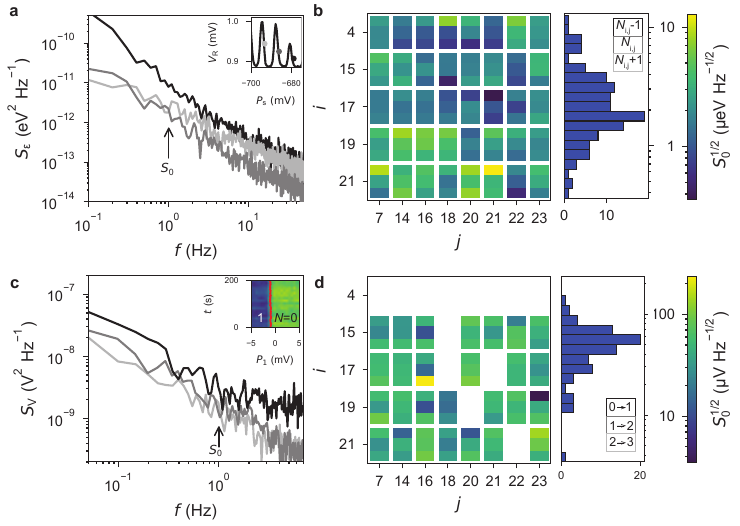}
\caption{\textbf{Charge noise characterisation}. \textbf{a,} The power spectral densities $S_\epsilon$ from the three measured Coulomb flanks of the charge sensor in tile $(4, 16)$. The spectra are fitted to a $S_0/f^{\gamma}$ dependence (Supplementary Fig.\.10). The inset shows the sensor reflectance at ohmic $O_1$ with respect to the sensor plunger $P_s$, showing three Coulomb peaks. with black, grey and light grey circles positioned at their flanks. \textbf{b,} Heat map of $S_0^{1/2}$ from the 40 investigated tiles (row $i$, column $j$) within the QARPET device. Each tile in the heat map is partitioned in three to report $S_0^{1/2}$ for increasing charge occupancy $N_{i,j}-1,N_{i,j},N_{i,j}+1$, corresponding to measurements from subsequent Coulomb flanks. The accompanying histogram shows the overall distribution, made of the total 120 experimental $S_0^{1/2}$ values. \textbf{c,} The voltage power spectral density of dot $D_1$ in tile $(4, 16)$ in the few hole regime calculated from tracking the transition voltages to charge occupancy $N=1,2,3$ (Methods). The spectra are fitted to a $S_0/f^{\gamma}$ dependence (Supplementary Fig.\.14). The inset shows a \SI{200}{s} repeated loading of the first hole in $D_1$ by sweeping the plunger $P_1$. The red line shows the estimated position for the $N=0$ to $N=1$ charge state transition. The transition voltage is then converted to the spectral density (Methods) shown in the panel. \textbf{d,} Similar to \textbf{b}, the heat map of the few hole regime $S_0^{1/2}$ corresponding to the physical location of each tile within the QARPET device.  A faulty RF line prevents fast measurements of tiles in row 4.  (Methods).}
\label{fig4}
\end{figure}

We characterize the charge noise properties of the sensors in the multi-hole regime using the flank method~\cite{connors_low_2019, lodari_low_2021,paquelet_wuetz_reducing_2023,massai_impact_2024}. 
Figure~\ref{fig4}a shows the obtained charge noise power spectral density $S_\epsilon$ as a function of frequency $f$ (Methods), measured at the flank of three neighbouring Coulomb peaks (inset Fig.\,\ref{fig4}a) of charge occupation $(n-1)$, $n$, $(n+1)$ to build up statistics.
We fit each spectrum with the function $S_0/f^{\gamma}$ to obtain the charge noise $S_\mathrm{0}^{1/2}$ at \SI{1}{Hz} (black arrow in Fig.\,\ref{fig4}a) and the spectral exponent $\gamma$, and iterate this protocol for all tiles under investigation (see Supplementary Figs.\,9,10).
The average $\gamma$ is close to 1 and is not correlated to $S_\mathrm{0}^{1/2}$ (see Supplementary Fig.\,11), suggesting that the observed $1/f$ trend results from an ensemble of two-level fluctuators with a wide range of activation energies~\cite{paladino20141, elsayed2024low}.

For each addressed tile $(i, j)$, the heat map in Fig.~\ref{fig4}b displays the charge noise $S_0^{1/2}$ measured at the three subsequent charge occupations, with the specific charge occupation affecting marginally the noise properties of the device (Supplementary Fig.\,12).  The overall distribution is shown in the accompanying histogram and is characterized by a spread over more than an order of magnitude, with a geometric mean of $2.4\pm 1.7$~\unit{\micro\electronvolt\,\hertz^{-1/2}}. This experimental mean matches closely the bootstrapped mean (Supplementary Fig.\,13), indicating that our sample size is sufficiently large to provide confidence in the accuracy of the mean value.
From the heat map, we identify a noisier device region in the bottom-left quadrant, $i \in \{19, 21\}, \; j \in \{14,  16\}$, as well as the best-performing tile, indexed $(17, 21)$, with an average charge noise $S_0^{1/2}$ of $0.7 \pm 0.24$~\unit{\micro\electronvolt\,\hertz^{-1/2}} and a minimum of \SI{0.36}{\micro\electronvolt\,\hertz^{-1/2}}. 

We complement the charge noise analysis by characterizing quantum dot $D_1$ under $P_1$ in the few-hole regime, which is the typical regime of operation for qubits.
Figure~\ref{fig4}c shows the frequency dependence of the voltage spectral density for the first three hole transitions. Following the methodology in Ref.~\cite{stehouwer_exploiting_2025}, these spectra are evaluated from the transition voltages time traces (inset Fig.\,\ref{fig4}c) obtained by sweeping across the charge transition region over a fixed time.
As in Fig.~\ref{fig4}a, we fit each spectrum to $S_0/f^{\gamma}$ to obtain the charge noise $S_0^{1/2}$ at \SI{1}{Hz} (black arrow in Fig.\,\ref{fig4}c) and the spectral exponent $\gamma$, and extend this protocol to all measurable tiles (Supplementary Fig\,14).
The resulting heat map in Fig.~\ref{fig4}d displays the charge noise values $S_0^{1/2}$ measured at the three subsequent charge transitions with the accompanying histogram showing the overall distribution. The distribution is characterized by a geometric mean value $S_0^{1/2}$ of $52 \pm 31$ \unit{\micro\volt\,\hertz^{-1/2}}. The charge noise average value does not change significantly with respect to hole filling, but rather the standard deviation decreases with increasing hole occupancy (Supplementary Fig.\,12).

Our statistical characterisation enables a comparison with charge noise measurements in quantum dot devices using a higher-quality Ge/SiGe heterostructure grown on a Ge substrate~\cite{stehouwer_exploiting_2025}.
We observe charge noise values an order of magnitude higher than in Ref.~\cite{stehouwer_exploiting_2025} for both charge sensors (Supplementary Fig.\,15) in the multi-hole regime and dots in the few-hole regime.
These results support the understanding from Ref.~\cite{stehouwer_exploiting_2025} that charge noise in Ge/SiGe heterostructures grown on Ge wafers may be lower than in those grown on Si wafers. Fabrication of QARPET devices on Ge/SiGe heterostructures grown on Ge wafers will improve statistical assessments of charge noise uniformity over larger areas.

\subsection{Qubits}\label{subsec5}
\begin{figure}
\centering
\includegraphics[width=0.5\textwidth]{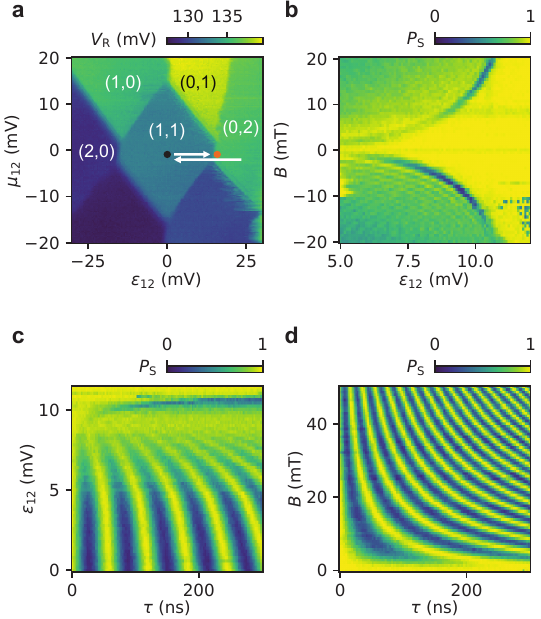}
\caption{ \textbf{Spin readout and singlet-triplet qubit.}
    \textbf{a,} Charge stability diagram of dot~1 and dot~2 measured on tile (19,14) from a second QARPET device, showing the sensor reflectance $V_\mathrm{R}$ as a function of detuning $\varepsilon_{12}$ and chemical potential $\mu_{12}$ of the double quantum dot. The red dot indicates the readout point inside the Pauli spin blockade window, the black dot the manipulation point, while the white arrows represent the pulse sequence from initialization to manipulation and readout. The gate voltages at the centre of the (1,1) region correspond to $P_\mathrm{s}$ = \SI{-1169.6}{mV}, $B_\mathrm{s}$ = \SI{-501.9}{mV}, $P_1$ = \SI{-663.6}{mV}, $P_2$ = \SI{-688.7}{mV}, $B_1$ = \SI{-385.0}{mV}, $B_2$ = \SI{-320.0}{mV}, ${S}$ = \SI{-146.5}{mV}.
    \textbf{b,} Experimental results for spin funnel, showing the singlet probability $P_\mathrm{S}$ as a function of $\varepsilon_{12}$ and applied magnetic field $B$. The applied magnetic field is nominally parallel to the device plane. The system is initialized in a singlet, pulsed from the (0,2) region towards the (1,1) region in \SI{1}{\micro s} at varying $\varepsilon_{12}$, left evolving for \SI{100}{ns} and pulsed to the readout point in \SI{1}{ns}. 
    \textbf{c} Experimental results for ST$_0$ oscillations, showing  $P_\mathrm{S}$ as a function of $\varepsilon_{12}$ and varying time $\tau$. At $B$ = \SI{20}{mT} the system is initialized in a singlet, pulsed from the (0,2) region towards the (1,1) region in \SI{1}{ns}, left evolving for varying time $\tau$, and pulsed to the readout point in \SI{1}{ns}. This sequence is repeated for different $\varepsilon_{12}$.
    \textbf{d} Experimental results for ST$_0$ oscillations, showing $P_\mathrm{S}$ as a function of $B$ and $\tau$. The system is is initialized in a singlet, pulsed from the (0,2) to the manipulation point in \SI{1}{ns} and left evolving for varying time $\tau$ and pulsed to the readout point in \SI{1}{ns}. This sequence is repeated for different magnetic fields.} 
\label{fig5}
\end{figure}

Lastly, we focus on a single tile of a second QARPET device with functional inter-dot barrier and demonstrate the ability to encode singlet-triplet (ST) and Loss-DiVincenzo (LD) spin qubits in the crossbar.
Figure \ref{fig5}a shows the charge stability diagram of the double dot $(D_1,D_2)$ obtained by sweeping the detuning $\varepsilon_{12}$ and constant potential axes $\mu_{12}$.
For ST qubits, we define the detuning of the two dots $\varepsilon_{12}$ to be zero at the centre of the (1,1) charge region and perform a spin funnel experiment. Starting from the (0,2) region, we pulse the system towards the (1,1) region at varying detuning $\varepsilon_{12}$, we let it evolve for \SI{100}{ns}, and pulse back to the readout point (red dot in Fig.~\ref{fig5}a). We do this for varying magnetic fields ($B$) to map the S-T$_{-}$ anti-crossing as a function of $B$ (dark blue line)~\cite{jirovec_singlet-triplet_2021}, which confirms the ability to read out spins using the Pauli spin blockade (PSB) method. 

Figure~\ref{fig5}c,d demonstrate coherent ST$_0$ oscillations as a function of detuning, magnetic field, and evolution time. We initialize a singlet in the (0,2) configuration, diabatically pulse towards (1,1), let evolve for a time $\tau$, and diabatically pulse to the readout point. This sequence is repeated for different $\tau$, for different detuning (Fig.~\ref{fig5}c) and different magnetic field (Fig.~\ref{fig5}d). From fitting the oscillation frequency as a function of magnetic field as shown in Supplementary Fig.\,15, we estimate the g-factor difference $\Delta g =0.0734 \pm 0.0001$ and the residual exchange at zero detuning $J(\varepsilon_{12}=0)=3.736 \pm 0.112$\SI{}{MHz}. 

\begin{figure}
\centering
\includegraphics[width=1\textwidth]{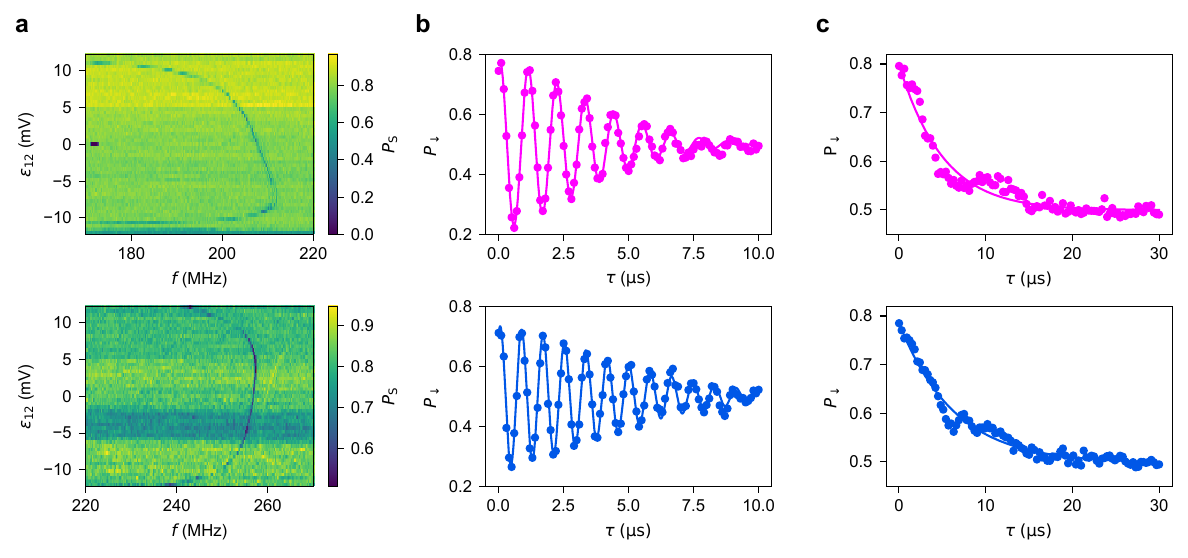}
\caption{
        \textbf{Coherence of Loss-DiVincenzo hole spin-qubits.}
        \textbf{a,} EDSR spectra of qubits $Q_1$ and $Q_2$, showing the singlet probability $P_\mathrm{S}$ as a function of detuning ($\epsilon_{12}$) and microwave pulse frequency $f$. Starting from the (0,2) configuration, we pulse in \SI{10}{\micro s} towards the (1,1) at varying detuning $\varepsilon_{12}$, apply a microwave pulse with frequency $f$ and pulse diabatically to the readout point. At zero detuning the $g$-factor of the two qubits are $g_1=0.36$ and $g_2=0.30$. A higher value in the colour bar represents a higher singlet probability.
        \textbf{b,} Ramsey experiment, showing the spin-down probability $P_\downarrow$ as a function of the waiting time $\tau$ between two $\pi/2$ pulses. Fitting of the results with $ P = a \cos(2\pi \Delta f \tau + \phi) \exp(-(\tau/T_2^*)^{\alpha^*})$, yields a $T_2^*$ of \SI[separate-uncertainty = true]{5.75(0.19)}{\micro s} and \SI[separate-uncertainty = true]{4.43(0.13)}{\micro s} for $Q_1$ and $Q_2$ respectively. 
        \textbf{c,} Hahn-echo experiment, showing the spin-down probability $P_\downarrow$ as a function of waiting time $\tau$. The pulse sequence consists of $\pi/2$, $\pi$, and $\pi/2$ pulses, separated by waiting time $\tau$.  Fitting the results with $P = a \exp(-\left({2\tau}/{T_2^H}\right)^{\alpha^H})$ yields a $T_2^H$ of \SI[separate-uncertainty = true]{12.69(0.40)}{\micro s} and \SI[separate-uncertainty = true]{10.11(0.42)}{\micro s} for $Q_1$ and $Q_2$, respectively. Initialization is performed pulsing deep in the (0,2) and waiting for \SI{100}{\micro s} for the system to relax into the singlet.
        All measurements are performed at a magnetic field of \SI{50}{mT}, oriented parallel to the device.
        }
\label{fig6}
\end{figure}

We measure the electric dipole spin resonance (EDSR) spectra of the two single hole spin LD qubits $Q_1$ and $Q_2$ (Fig.\,\ref{fig6}a) by starting with a singlet in the (0,2) configuration, pulsing slowly towards the (1,1), applying a microwave (MW) pulse to plunger $P_1$ / $P_2$ (top/bottom panel) at frequency $f$, and pulsing to the readout point. A dark blue line is visible where the MW driving frequency is at resonance with the qubit. At zero detuning, that is at the centre of the (1,1) region, the $g$-factors for the two qubits result in $g_1=0.30$ and $g_2=0.36$ (in line with the values reported for the same heterostructure in Ref.~\cite{hendrickx_single-hole_2020} and with other works on Ge~\cite{jirovec_dynamics_2022, hendrickx2024sweet}). The difference of these $g$-factors is also consistent with the $\Delta g=0.073$ extracted from the ST qubit experiment in Fig.~\ref{fig5}b.

After calibrating the $\pi/2$ qubit rotations we characterize the qubits' coherence in this system. We perform a Ramsey experiment (Fig.\,\ref{fig6}b) on both qubits at a magnetic field of \SI{50}{mT} and extract coherence time $T_2^*$ of \SI[separate-uncertainty = true]{4.43(0.13)}{\micro s} and \SI[separate-uncertainty = true]{5.75(0.19)}{\micro s} for $Q_1$ and $Q_2$ respectively. These coherence times are comparable to the best results measured in Ge at a similar magnetic field \cite{Wang2024OperatingSpins}. Further we perform a Hahn-echo experiment (Fig.\,\ref{fig6}c) and extract coherence times $T_2^H$ of \SI[separate-uncertainty = true]{10.11(0.42)}{\micro s} and \SI[separate-uncertainty = true]{12.69(0.40)}{\micro s} for $Q_1$ and $Q_2$ respectively, comparable with previous experiments~\cite{Wang2024OperatingSpins} (note that in Ref.\cite{Wang2024OperatingSpins} $T_2^H$ of \SI{32}{\micro s} is measured at at \SI{25}{mT} where coherence is expected to be longer compared to our field of \SI{50}{mT}). This proof of principle demonstration further motivates the use of QARPET devices for statistical characterisation of qubit coherence and is encouraging for scaling quantum processors in dense Ge qubit arrays.

\section{Conclusions}\label{sec3}

We have demonstrated key functionalities of the QARPET architecture at millikelvin temperature using RF reflectometry, including tile selectivity, single-hole quantum dots, singlet-triplet, and single-spin Loss-DiVincenzo qubits. We gathered statistics on single-hole and addition voltages, providing insights into the uniformity of gate-defined quantum confinement in our device architecture, relevant for hole-spin qubit operations~\cite{Bosco2021SqueezedPower, Malkoc2016OptimalQubits} and for developing shared-control spin qubit architectures~\cite{borsoi_shared_2024}. The systematic exploration of charge noise showcases the power of the architecture as a measurement tool for statistical benchmarking and motivates further fabrication and characterisation of similar devices on improved heterostructures grown on germanium wafers. Together with the proof of principle demonstration of spin qubits with respectable quantum coherence, these results mark a significant advance towards the statistical study of spin qubits within a single cooldown and across µm-scale length-scales, which are relevant for noisy intermediate scale quantum processors. The integration of autonomous, machine--learning assisted routines for efficient tuning, readout, and control of spin qubits~\cite{carlsson_automated_2025} will be a key step in this direction.

QARPET is a flexible platform with potential for further optimization in both design and implementation. While the current design features two quantum dots per tile for charge and spin manipulation, this number could be extended to $q > 2$, requiring a linear increase in control lines $(n+m+2q+3)$ to operate $q \times n \times m$ qubits, thereby maintaining a sublinear scaling of interconnects. The specific properties of the heterostructure, such as the SiGe barrier thickness, also influence the dimensions of the plunger and barrier gates, thereby setting the minimum tile footprint and determining the overall qubit density of the device.

Although demonstrated in a Ge/SiGe heterostructure, the QARPET architecture could be adapted to other accumulation-mode undoped heterostructures with appropriate design and process modifications. However, such adaptations may introduce additional challenges. For instance, implementing QARPET in Si/SiGe heterostructures would require narrower gates due to the the heavier electron mass compared to holes in strained Ge, as well as additional accumulation gate layers to connect charge sensors to remote doped reservoirs. 

With hundreds of nominally identical quantum dot qubits integrated in a single die, we anticipate the use of QARPET as an ideal test-bed for training of automated routines for all phases of the device tuning, from quantum dot read out, to single-spin operation, also exploiting machine learning and artificial intelligence~\cite{zwolak_data_2024,alexeev_artificial_2024}. On the other hand, implementing test vehicles like QARPET with advanced semiconductor manufacturing will reduce device variability stemming from the material stack and fabrication, and accelerate the development-cycle of industrial spin qubits. 
Additionally, due to the dense pattern of multi-layer gates, QARPET-like devices may provide benchmarking of spin qubit performance in a practically relevant electrostatic environment, paving the way for the development of large-scale quantum processors.

\section{Methods}\label{sec4}

\textbf{Ge/SiGe heterostructure growth.} The Ge/SiGe heterostructure is grown on a 100-mm n-type Si(001) substrate using an Epsilon 2000 (ASMI) reduced pressure chemical vapor deposition reactor. The layer sequence comprises a $\mathrm{Si_{0.2}Ge_{0.8}}$ virtual substrate obtained by reverse grading, a \SI{16}{nm} thick Ge quantum well, a \SI{55}{nm}-thick $\mathrm{Si_{0.2}Ge_{0.8}}$ barrier, and a thin sacrificial Si cap. Further details and electrical characterisation of Hall-bar shaped heterostructure field effect transistors on this semiconductor stack are presented in Ref.~\cite{lodari_low_2021}.

\textbf{Device fabrication.} The fabrication of QARPET devices entails the following steps: Electron-beam lithography of the ohmic contacts layer; Wet etching of the sacrificial Si-cap in buffer oxide etch for \SI{10}{s}; Deposition of the ohmic contacts via e-gun evaporation of \SI{15}{nm} of Pt at pressure of \SI{3e-6}{mbar} at the rate of \SI{3}{nm/minute}, followed by rapid thermal anneal  at \SI{400}{C} for \SI{15}{minutes} in a halogen lamps heated chamber in argon atmosphere to form PtSiGe ohmic contacts~\cite{tosato_hard_2023}; Atomic layer deposition of \SI{5}{nm} of $\mathrm{Al_2O_3}$ at \SI{300}{C}; Electron-beam lithography and deposition of the first gate layer via e-gun evaporation of \SI{3}{nm} of Ti and \SI{17}{nm} of Pd; Lift-off in AR600-71 \SI{45}{C} with sonication at medium-high power for 1 hour, the patterned side of the chip facing downwards to avoid re-deposition of metal on the chip surface. For each subsequent gate layer the last two steps are repeated, increasing each time the deposited Pd thickness by \SI{5}{nm} to guarantee film continuity where overlapping with the first gate layer. We perform AFM imaging of the developed resist after each resist development and metal lift-off to monitor the fabrication process.

\textbf{Measurements} 
Device 1 was screened for leakage at 4.2 K before mK measurements, yield details are in Supplementary Figure 1. Possible failure modes for this architecture are described in Supplementary Note 1.
All measurements are performed in a Oxford wet dilution refrigerator with a base temperature of $\approx$ 100 mK. Using battery‐powered voltage sources, dc‐voltages are applied to the gates. The DC voltages on gates are combined with an AC voltage from a Qblox arbitrary waveform generator (AWG)  by a bias‐tee with a cut‐off frequency of \SI{3}{Hz}. The AC‐voltage used for pulses and RF driving is generated by an AWG .  We studied two devices. Measurements and corresponding analysis reported in  Figs.~2-4 and Supplementary Figs.\,1--15. are from QARPET device 1. The inter-dot barrier $B_2$ was leaky, preventing double-dot studies on this specific device and the RF line connected to  for $i=4$ was faulty, preventing on this row to perform fast 2D maps of $P_1$ vs other gates. In QARPET device 2 we focused on a single tile with measurements and corresponding analysis reported in Figs~5,6 and Supplementary Fig.\,16. The magnetic field is applied via a one-axis solenoid magnet that is nominally parallel to the device plane. 

\textbf{Relative lever arm calculation} 
The relative lever arms to $D_1$ with respect to $P_1$ in Figure~\ref{fig3}a are derived from the slopes of the transition lines in the reflectance maps of all $P_1$-to-surrounding-gates pairs~\cite{Tidjani2023VerticalWell} for all the tiles where it was measurable (Supplementary Figs.\,4--7).
The slope of the transition lines in these gate-$P_1$ maps allows to extract the lever arm of gate $g_i$ to $D_1$ relative to the lever arm of $P_1$ to $D_1$  ($\alpha_{g_i,D_1}/\alpha_{P_1,D_1}$)~\cite{Tidjani2023VerticalWell}. The variability of the relative lever arms across different spin qubit tiles arises form differences in the shape and position of dot $D_1$, which depends on the electrostatic potential surrounding the dot and reflects the variability in semiconductor stack and gate stack uniformity, as well as differences in electrostatic tuning of the tile.
The virtual gate voltage $\text{v}P1$ is calculated as $\text{u}P_1 \cdot \text{median}(P_1)/\text{median}(\text{u}P_1)$ where $\text{u}P_1 = \sum g_i \alpha_{g_i,D_1}/\alpha_{P_1,D_1}$ with $g_i$ comprising $P_s$, $B_s$, $B_1$, $P_1$ and $S$.

\textbf{Flank method for charge noise characterisation of the quantum dots in the multi-hole regime}
We characterize charge noise of the sensors using the flank method, measured via RF reflectometry. The process begins by tuning the surrounding gates until the reflected signal exceeds the baseline by at least 10 mV, marking the turn-on of the device. A multi-hole quantum dot is then defined beneath $P_s$ by gradually lowering the barrier gates surrounding $P_s$ until a spectrum of Coulomb peaks is observed without any background signal.
Next, we record the reflected signal on the right flank of the first three Coulomb peaks, where the slope ($|dI/dV_{sd}|$) of the peaks is steepest. The signal is sampled at a rate of 1 kHz for a duration of 100 seconds using a Qblox digitizer. To compute the current power spectral density $S_I$, the 100-second current trace is divided into ten 10-second segments. The power spectral densities for each segment are calculated and averaged to yield $S_I$.
For each Coulomb peak analysed, we convert $S_I$ into the charge noise power spectral density ($S_\epsilon$) using:
\begin{equation}
    S_\epsilon = \frac{\alpha^2 S_I}{|dI/dV_{sd}|^2}
\end{equation}

where $\alpha$ is the lever arm extracted from the analysis of the Coulomb diamonds (Supplementary Fig.\,9). Finally, the spectral densities are fitted to a $1/f^\gamma$ model, and the value of the spectral density at 1 Hz ($S_0^{1/2}$) and exponent ($\gamma$) are extracted and reported.

\textbf{Charge transition method for charge noise characterisation of quantum dots in the few-hole regime}
We begin by tuning $P_1$ to the last-hole, as described previously. To probe electrostatic fluctuations in the quantum dot, we repeatedly sweep across the charge transition point over a period of \SI{200}{s} at a rate of 300 Hz. Transition voltages are extracted by performing a sigmoid fit to each sweep repetition~\cite{stehouwer_exploiting_2025}.
From the transition voltage data, we calculate the voltage spectral density and fit the results to a $1/f^\gamma$ dependence, reporting the noise value at 1 Hz and the spectral exponent. Because we use the sensor to monitor the transition voltage, the measurements include noise contributions from both $P_s$ and $P_1$. However, we find that the spectral density of the sensor is an order of magnitude lower than the spectral densities measured for $P_1$. This confirms that the observed traces are dominated by charge noise from the quantum dot under $P_1$~\cite{stehouwer_exploiting_2025}.

\backmatter

\section*{Supplementary Information}
Supplementary Figs.\,1--16 and Supplementary Note 1.

\section*{Acknowledgements}
We are grateful to Francesco Borsoi, Pablo Cova Fari\~{n}a, Hanifa Tidjani, Stephan Philips, and Daniel Jirovec for fruitful discussions.
We acknowledge support by the European Union through the IGNITE project with grant agreement No. 101069515 and the QLSI project with grant agreement No. 951852.
This work was supported by the Netherlands Organisation for Scientific Research (NWO/OCW), via the Open Competition Domain Science - M program.
This research was sponsored in part by the Army Research Office (ARO) under Awards No. W911NF-23-1-0110. The views, conclusions, and recommendations contained in this document are those of the authors and are not necessarily endorsed nor should they be interpreted as representing the official policies, either expressed or implied, of the Army Research Office (ARO) or the U.S. Government. The U.S. Government is authorized to reproduce and distribute reprints for Government purposes notwithstanding any copyright notation herein.
This research was sponsored in part by The Netherlands Ministry of Defence under Awards No. QuBits R23/009. The views, conclusions, and recommendations contained in this document are those of the authors and are not necessarily endorsed nor should they be interpreted as representing the official policies, either expressed or implied, of The Netherlands Ministry of Defence. The Netherlands Ministry of Defence is authorized to reproduce and distribute reprints for Government purposes notwithstanding any copyright notation herein.

\section*{Declarations}
\begin{itemize}
\item G.S. and A.T. are inventors on a patent application (International Application No. PCT/NL2024/050325) submitted by Delft University of Technology related to the QARPET architecture. G.S. is founding advisor of Groove Quantum BV and declares equity interests. 
\item The datasets supporting the findings of this study are openly
available at https://zenodo.org/records/15089359.
\item  A.T. designed and fabricated the devices on heterostructures provided by L.E.A.S, with contributions from D.C. to packaging, from K.H. to device inspection, and from D.D.E for process development. 
A.T. assembled the setup and fridge, performed quantum dot and qubit measurements, and analysed the corresponding data. A.E. and F.P. performed noise measurements, analysed with help from L.E.A.S. and D.D.E. 
A.T., A.E., and G.S. wrote the manuscript with input from all authors.
G.S. and A.T. conceived the project, supervised by G.S.
\end{itemize}


\end{document}


\title[Article Title]{QARPET: A Crossbar Chip for Benchmarking Semiconductor Spin Qubits (Supplementary Information)}

 \author[1]{\fnm{Alberto} \sur{Tosato}}
 \author[1]{\fnm{Asser} \sur{Elsayed}}
 \author[1]{\fnm{Federico} \sur{Poggiali}}
 \author[1]{\fnm{Lucas} \sur{Stehouwer}}
 \author[1]{\fnm{Davide} \sur{Costa}}
 \author[1]{\fnm{Karina} \sur{Hudson}}
 \author[1]{\fnm{Davide} \sur{Degli Esposti}}
 \author*[1]{\fnm{Giordano} \sur{Scappucci}}\email{g.scappucci@tudelft.nl}

 \affil[1]{\orgdiv{QuTech and Kavli Institute of Nanoscience}, \orgname{Delft University of Technology}, \orgaddress{\street{Lorentzweg 1}, \postcode{2628 CJ} \city{Delft}, \country{The Netherlands}}}

 \maketitle

\renewcommand{\figurename}{Supplementary Fig.}

\setcounter{table}{0}
\setcounter{figure}{0}

\newpage

\begin{figure*}[t]
    \centering
    \includegraphics[width=1\linewidth]{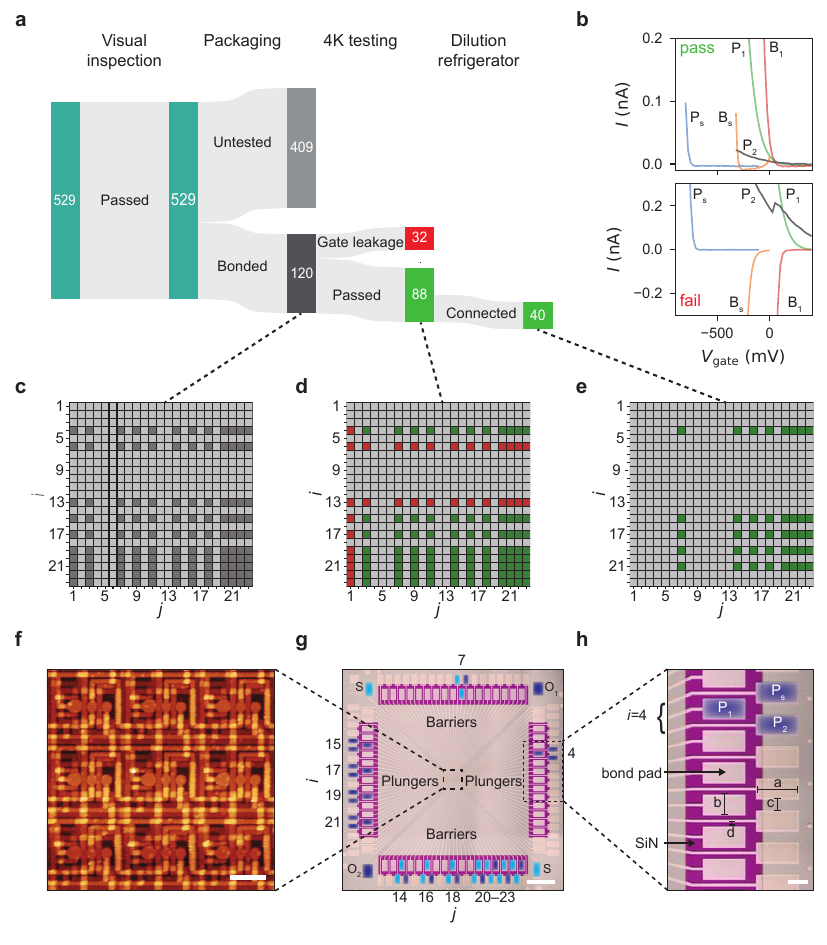}
    \caption{\textbf{Screening of QARPET device tiles.}  \textbf{a,} QARPET device 1 underwent a visual screening by microscopy as well as electrical transport screening at a temperature of $T$ = 4.2~K.
Out of 529 tiles fabricated, all passed inspection by optical and atomic force microscopy imaging. 
120 tiles were down-selected for bonding and electrical screening at a temperature of 4.2~K, prioritizing alignment feasibility between the bond-pads on the device and on the printed circuit board that is detailed in Ref.~\cite{philips_universal_2022}. 
Excluding the $B_\mathrm{2}$ gate, which was leaky and affected all tiles due to its shared bond, 88 tiles (73\% yield) passed the criterion requiring all other gates ($P_\mathrm{s}$, $B_\mathrm{s}$, $B_\mathrm{1}$, $P_\mathrm{1}$, $P_\mathrm{2}$) to accumulate or deplete the hole channel without leakage.
40 tiles were connected for millikelvin temperature measurements, due to the limited number of radio-frequency (RF) lines in the dilution refrigerator  and the intentional removal of bonds from leaky rows/columns as a precaution. 
\textbf{b,} Example of measured current $I$ vs. gate voltage $V_\textrm{{gate}}$ at 4.2~K, where each curve corresponds to a different gate  ($P_\mathrm{s}$, $B_\mathrm{s}$, $B_\mathrm{1}$, $P_\mathrm{1}$, $P_\mathrm{2}$.). Top: passing tile; bottom: failing tile.
\textbf{c,} Schematic of the $23\times23$ array of tiles (index row $i$, column $j$) with bonded tiles in dark grey and untested tiles in light grey.\textbf{d} Tiles in green passed leakage and transport testing at 4.2~K, tiles in red failed.
\textbf{e,} Tiles (green) connected for millikelvin measurements.
\textbf{f,} Representative atomic-force microscopy detail of a patch of nine tiles. Scale bar: 500 nm.
\textbf{g,} Representative optical image of of a QARPET chip before bonding. Scale bar: 0.5~mm. Bond-pads used for millikelvin measurements are boxed: cyan for DC lines, blue for DC+RF lines. Purple silicon nitride (SiN) strips insulate selected pads from the substrate for room-temperature leakage testing.
\textbf{h,} Detail of the alternating bond-pad layout showing plunger gate connections for row $i = 4$. Scale bar: $50~\upmu\textrm{m}$. Bond-pad width $a = 200~\upmu\textrm{m}$, height $b = 100~\upmu\textrm{m}$, separation in the outer row $c = 50~\upmu\textrm{m}$, and minimal distance to nearby fanout line in the inner row $d = 10~\upmu\textrm{m}$.}
    \label{fig:Sq}
\end{figure*}

\newpage
\begin{landscape}
\begin{figure}[h]
    \centering
    \includegraphics[width=1\linewidth]{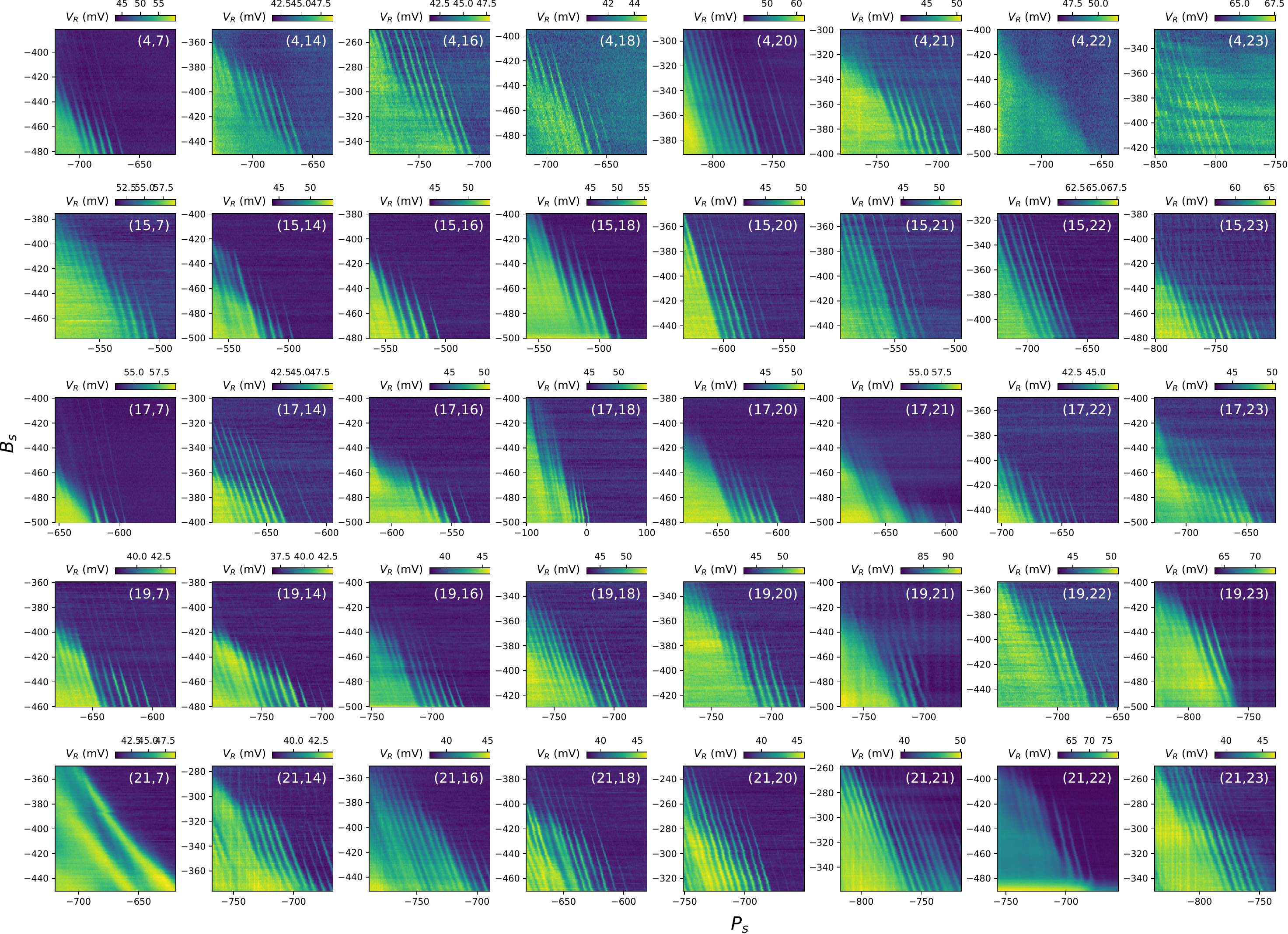}
    \label{fig:enter-label}
\caption{This figure extends Fig.~2a by including color bars and voltage range for all tiles. $V_\mathrm{R}$ is the sensor reflectance, $P_\mathrm{s}$ is the plunger sensor gate voltage, all axes are in units of mV.}
\end{figure}
\end{landscape}

\newpage

\begin{landscape}
    
\begin{figure}[h]
    \centering
    \includegraphics[width=1\linewidth]{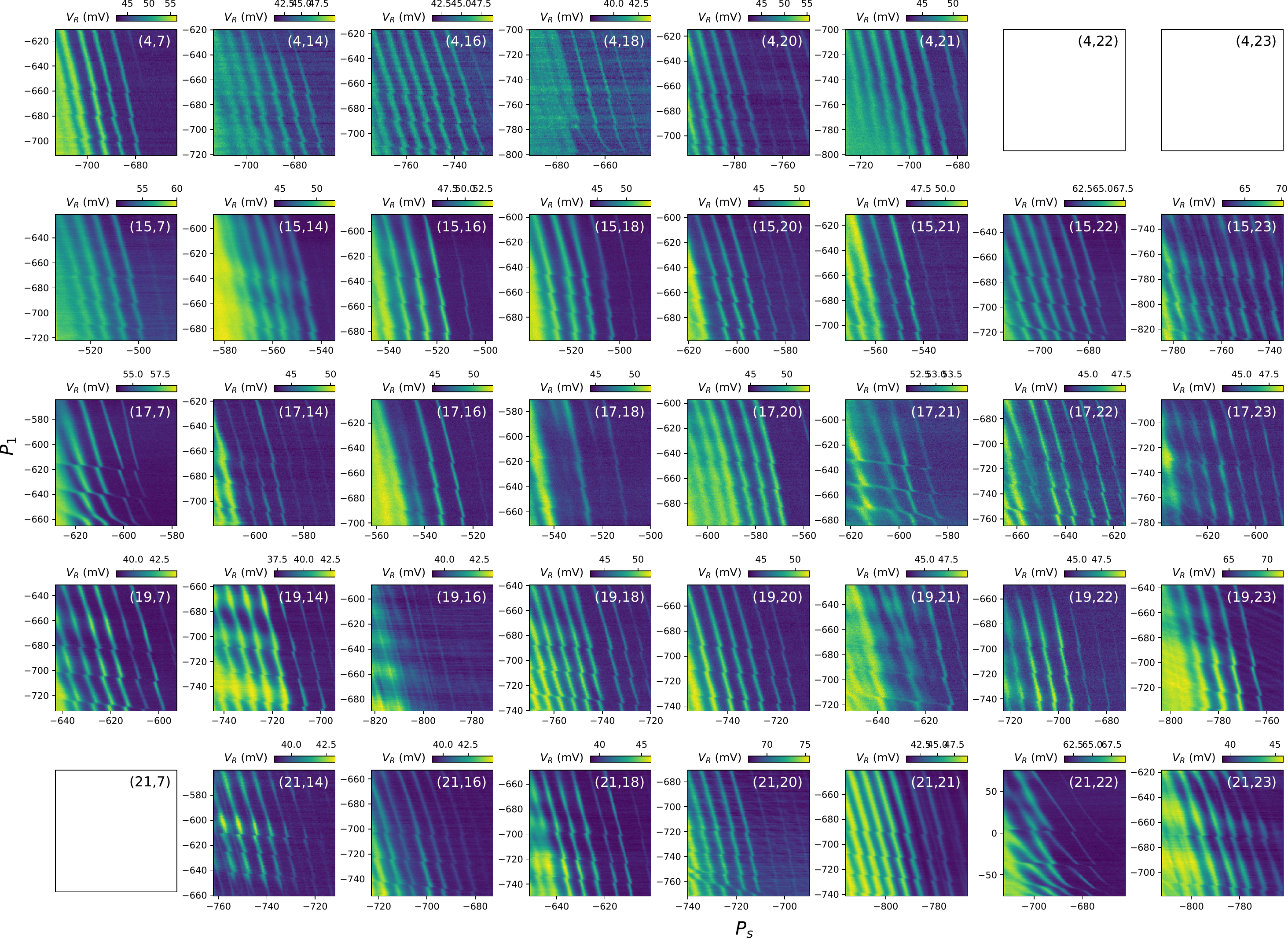}
    \label{fig:enter-label}
\caption{This figure extends Fig.~2b by including color bars and voltage range for all tiles. $V_R$ is the sensor reflectance, all axes are in units of mV.}
\end{figure}
\end{landscape}

\newpage
\begin{landscape}
    
\begin{figure*}[!ht]
    \centering
    \includegraphics[width=\linewidth]{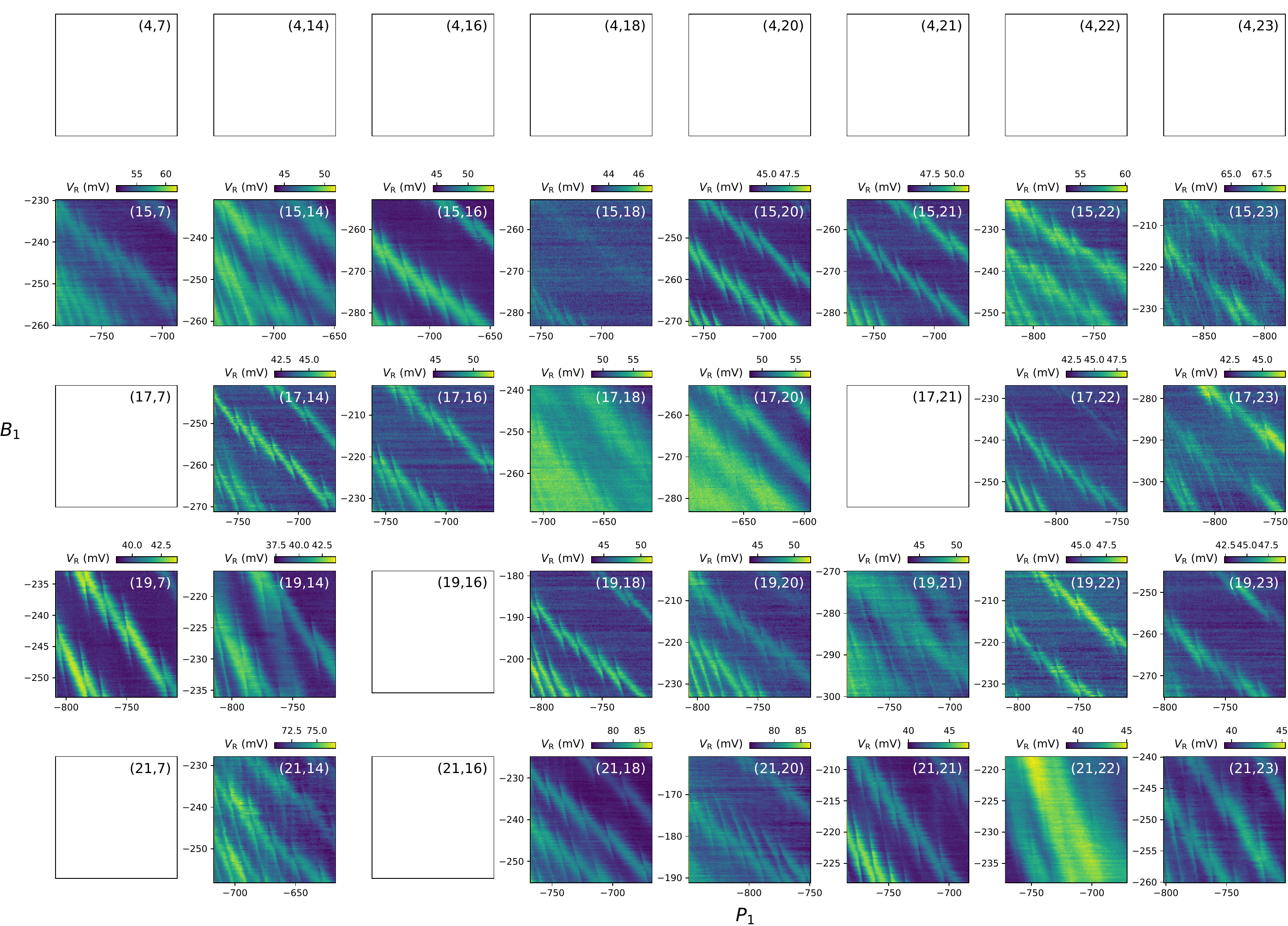}
    \caption{Reflectance maps of $B_1$ (y-axis) vs $P_1$ (x-axis) used to determine the relative lever arm values presented in Fig.~3a. The slope of the transitions is used to determine the relative lever arm to $D_1$ $\alpha_{B_1,D_1}/\alpha_{P_1,D_1}$. Measurement on row 4 were not performed as the RF line connected to $P_1$ for $i=4$ was faulty therefore preventing efficient measurements on those tiles. The remaining missing maps were not measured because the voltage range would violate the constraints sets for the device. The same considerations apply to the reflectance maps in Supplementary Figs.~5--7. $V_R$ is the sensor reflectance, all axes are in units of mV.}
    \label{figS2}
\end{figure*}
\end{landscape}
\newpage

\begin{landscape}
    
\begin{figure*}[!ht]
    \centering
    \includegraphics[width=\linewidth]{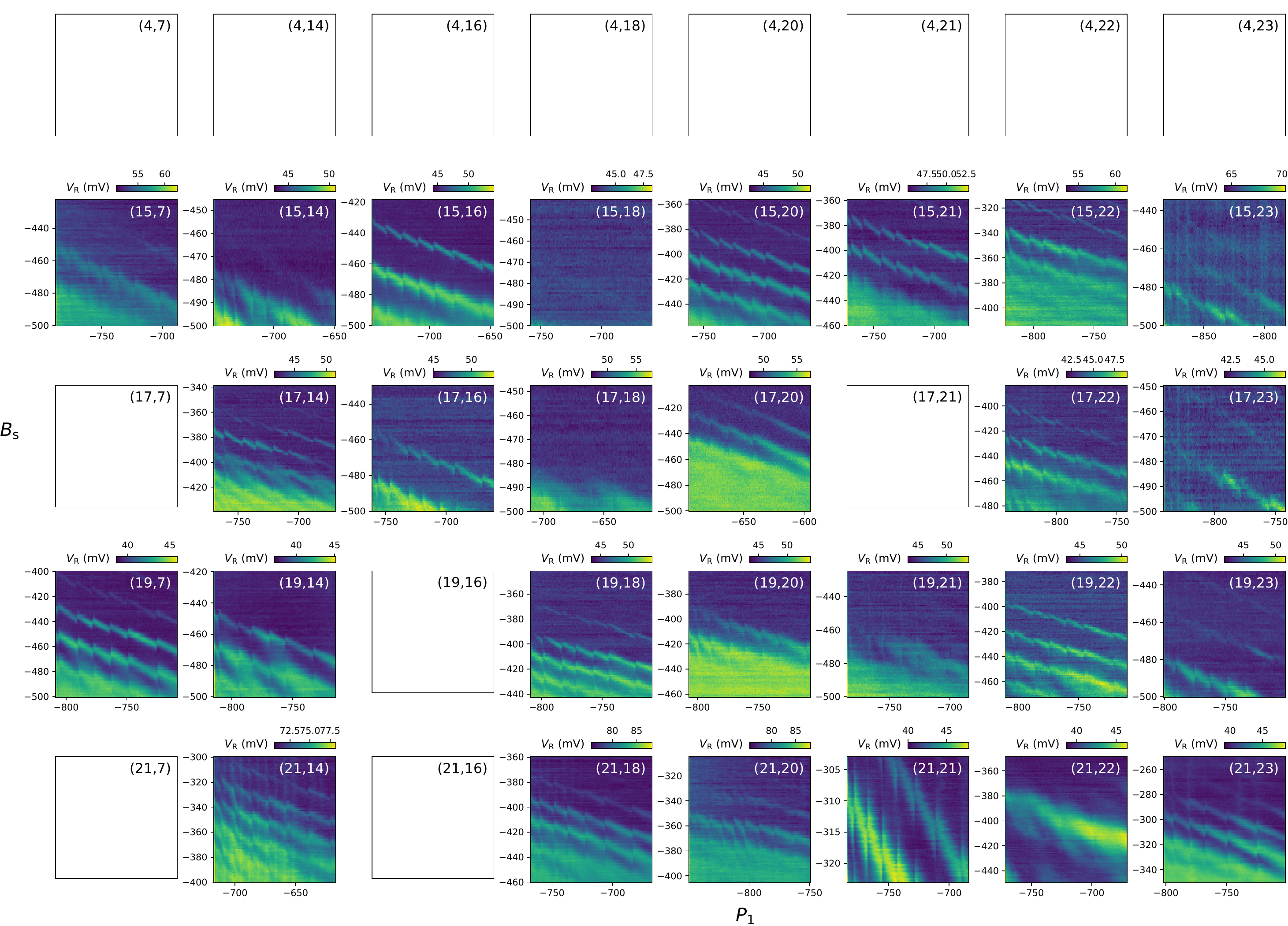}
    \caption{Reflectance maps of $B_s$ (y-axis) vs $P_1$ (x-axis) used to determine the relative lever arm values presented in Fig.~3a. The slope of the transitions is used to determine the relative lever arm to $D_1$ $\alpha_{B_s,D_1}/\alpha_{P_1,D_1}$. $V_R$ is the sensor reflectance, all axes are in units of mV.}
    \label{figS3}
\end{figure*}
\end{landscape}

\newpage
\begin{landscape}
    
\begin{figure*}[!ht]
    \centering
    \includegraphics[width=\linewidth]{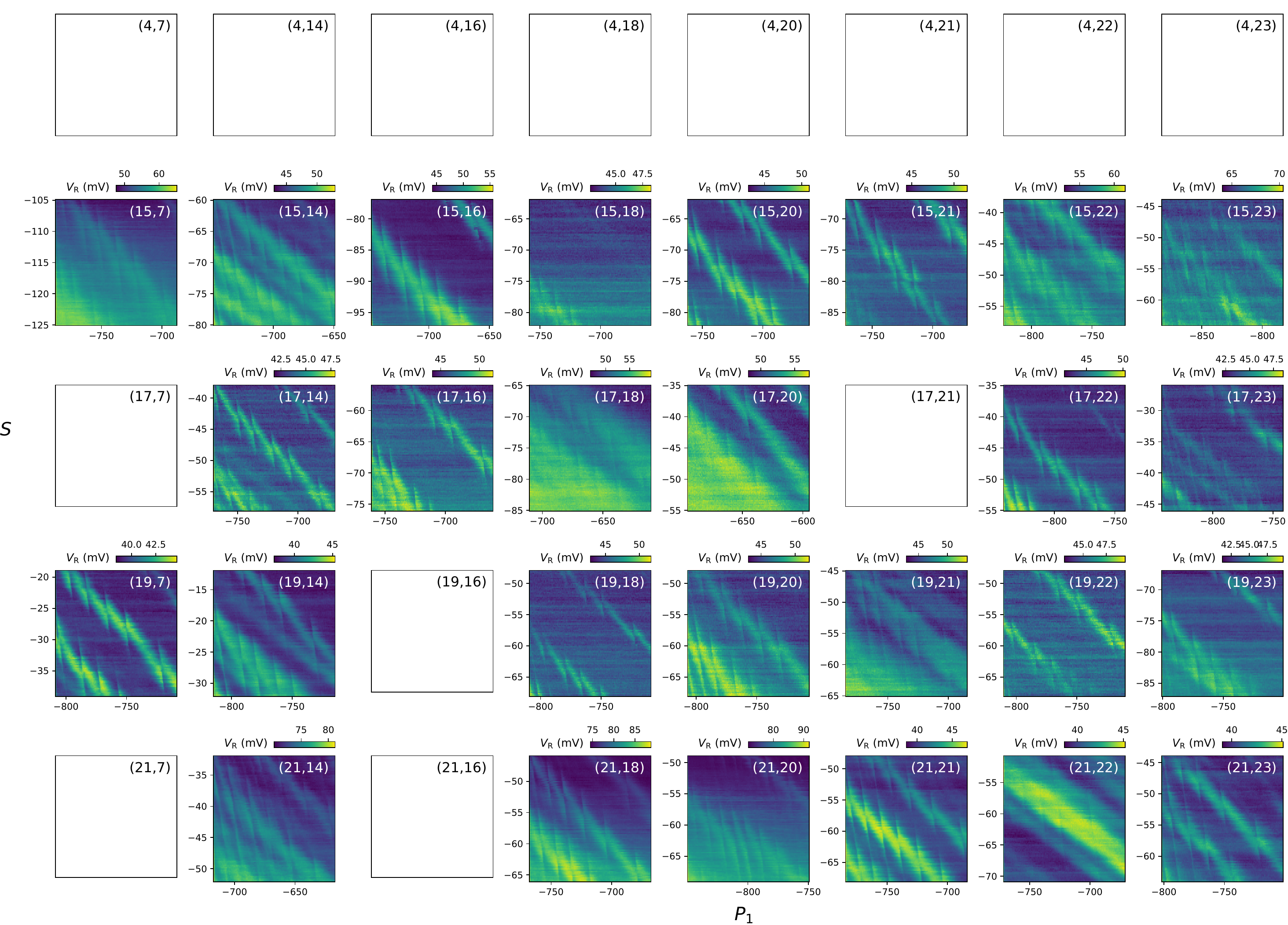}
    \caption{Reflectance maps of $S$ (y-axis) vs $P_1$ (x-axis) used to determine the relative lever arm values presented in Fig.~3a. The slope of the transitions is used to determine the relative lever arm to $D_1$ $\alpha_{B_s,D_1}/\alpha_{P_1,D_1}$. $V_R$ is the sensor reflectance, all axes are in units of mV.}
    \label{figS4}
\end{figure*}
\end{landscape}

\newpage
\begin{landscape}

\begin{figure*}[!ht]
    \centering
    \includegraphics[width=\linewidth]{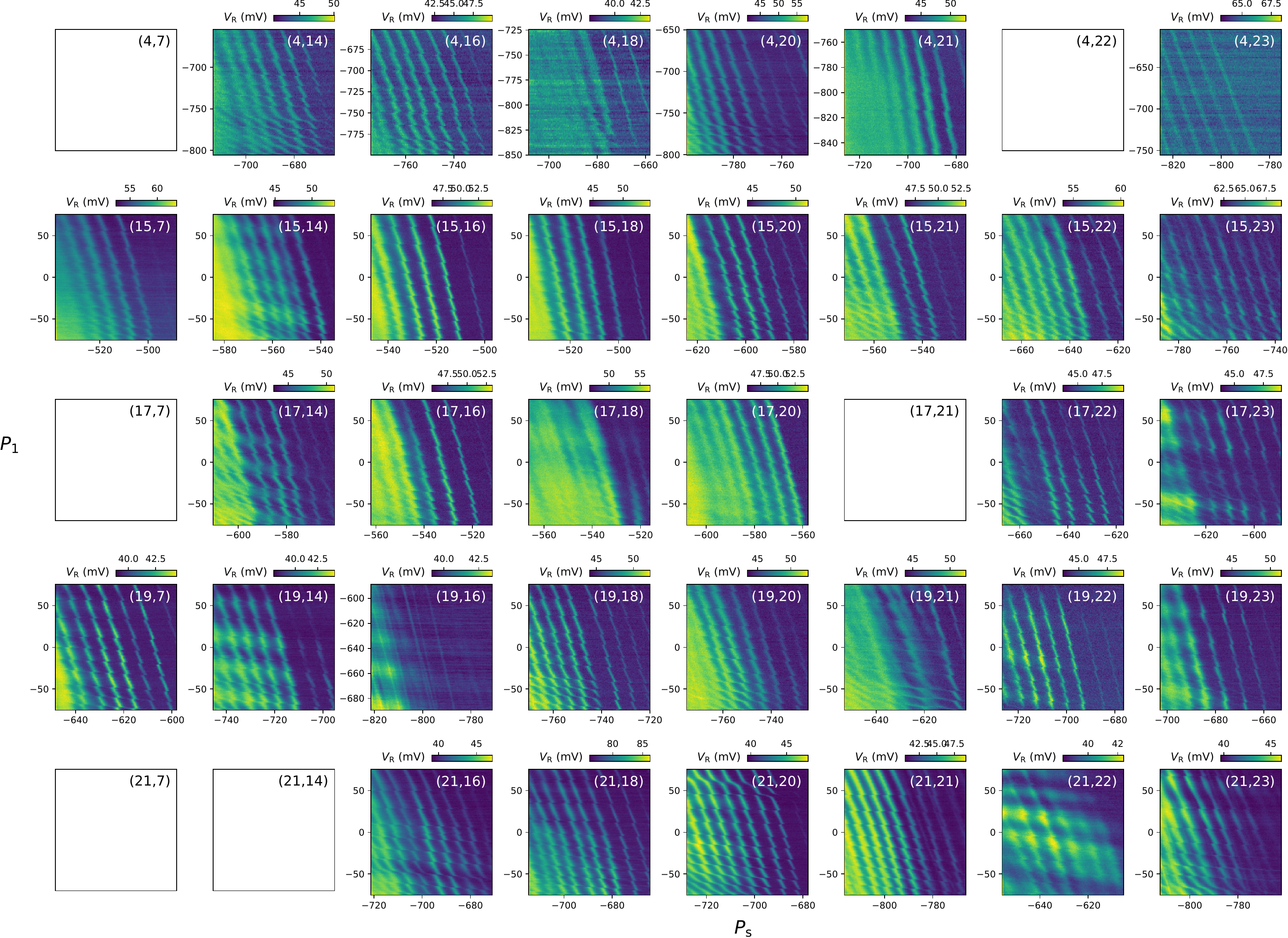}
    \caption{Reflectance maps of $P_1$ (y-axis) vs $P_s$ (x-axis) used to determine the charging voltages presented in Fig.~3c. The first transition line at the top corresponds to the last hole transition. $V_R$ is the sensor reflectance, all axes are in units of mV.}
    \label{S5}
\end{figure*}
\end{landscape}
\newpage
\begin{figure}[ht]
    \centering
    \includegraphics[width=0.5\linewidth]{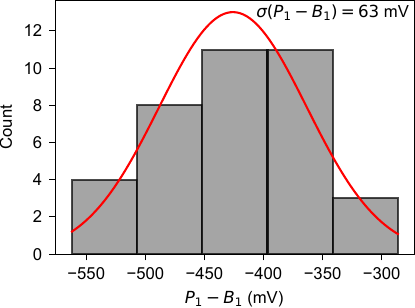}
    \caption{Histogram of voltage differences between plunger $P_1$ and barrier gates $B_1$ for the first hole transition across 37 tiles of device 1 with superimposed Gaussian (red line). $(P_1-B_1)$ is calculated from the distributions of $P_1$ and $B_1$ reported as violin plots in in Fig.~3b of the main text. The obtained standard deviation $\sigma(P_1-B_1)$ is 63~mV. Rigorous benchmarking to prior literature on silicon quantum dots \cite{ha_flexible_2022,Neyens2024ProbingWafers} is challenged by the differences in material systems (Ge \textit{vs.} Si, holes \textit{vs.} electrons), device design (gate shape and size), the ensemble of devices considered (number and their spacing), and the details of the measurement tuning conditions.}
    \label{fig:enter-label}
\end{figure}

\newpage
\begin{landscape}

\begin{figure*}[!ht]
    \centering
    \includegraphics[width=\linewidth]{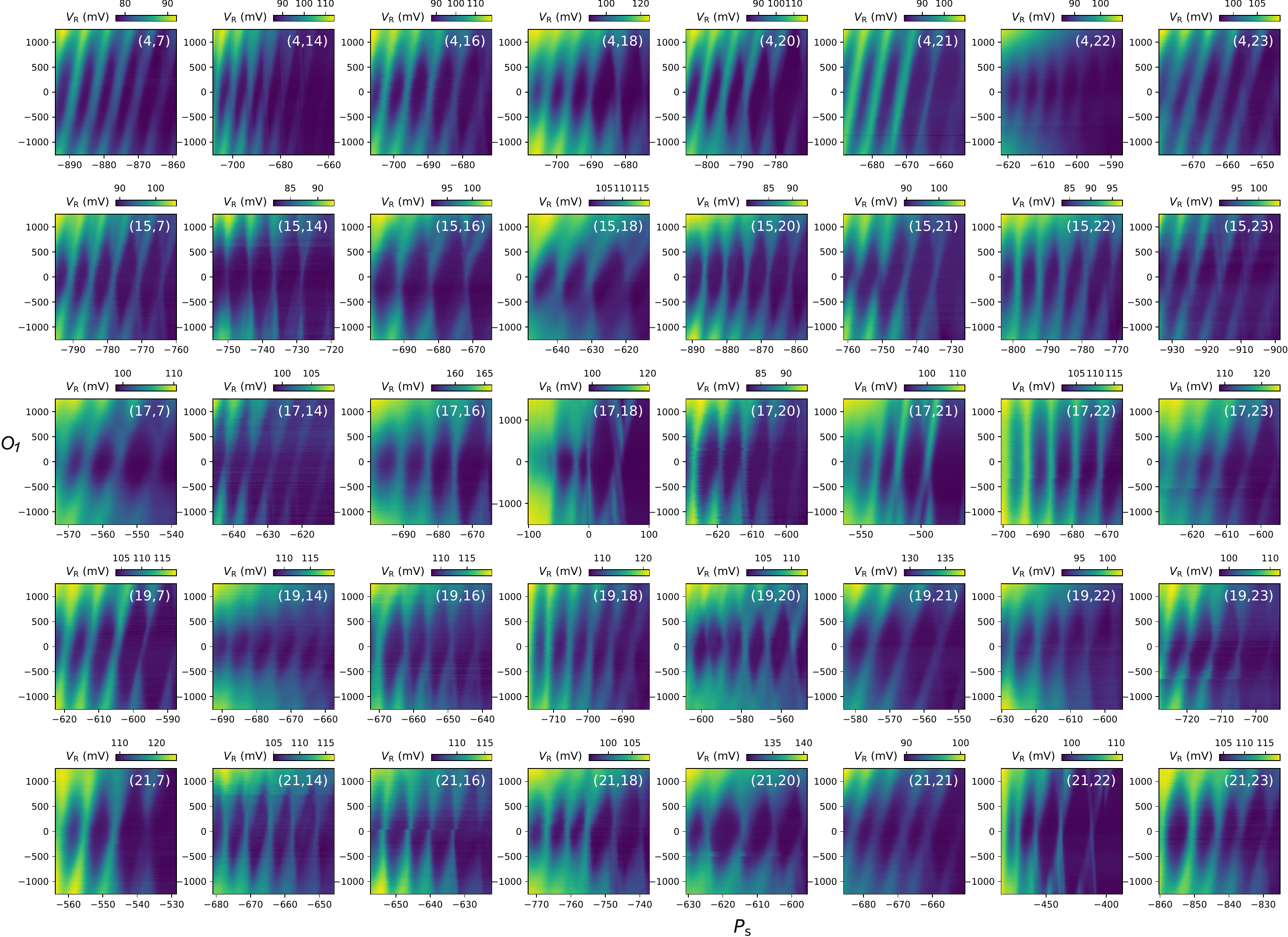}
    \caption{The reflectance measured as a function of sensor plunger ($P_s$) and voltage applied to ohmic contact ($O_{1}$) showing Coulomb diamonds for individual tiles.}
    \label{figS6}
\end{figure*}
\end{landscape}

\newpage
\begin{landscape}
\begin{figure*}[!ht]
    \centering
    \includegraphics[width=\linewidth]{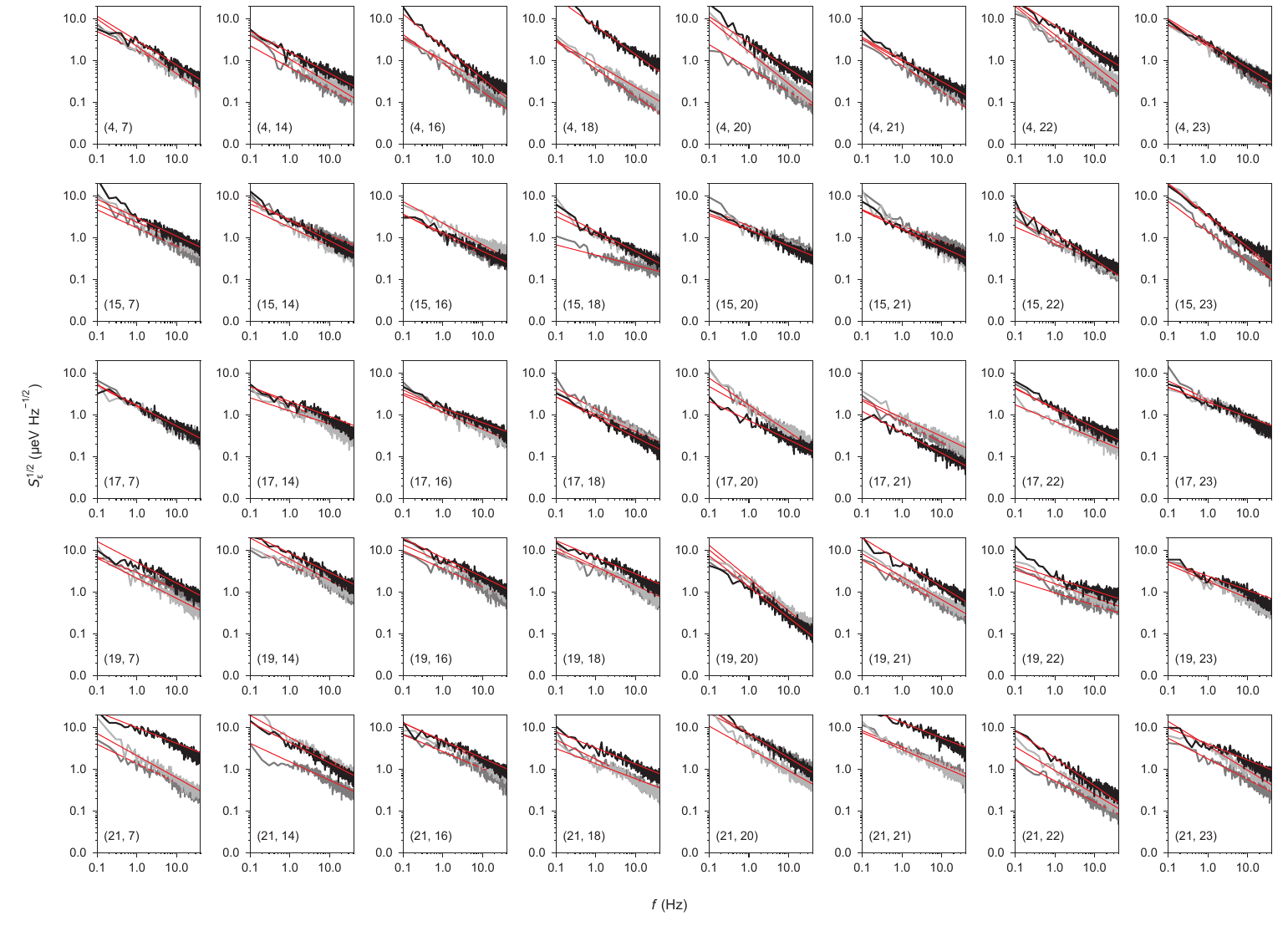}
    \caption{The power spectral densities $S_\epsilon^{1/2}$ from the three measured coulomb flanks of the sensor dot in each of the 40 investigated tiles. The tile index is indicated in each panel. The spectral densities are individually fit to a $1/f^\gamma$ and both $S_0$ and $\gamma$ are extracted.}
    \label{figS7}
\end{figure*}
    
\end{landscape}
\newpage

\begin{figure*}[!ht]
    \centering
    \includegraphics[width=\linewidth]{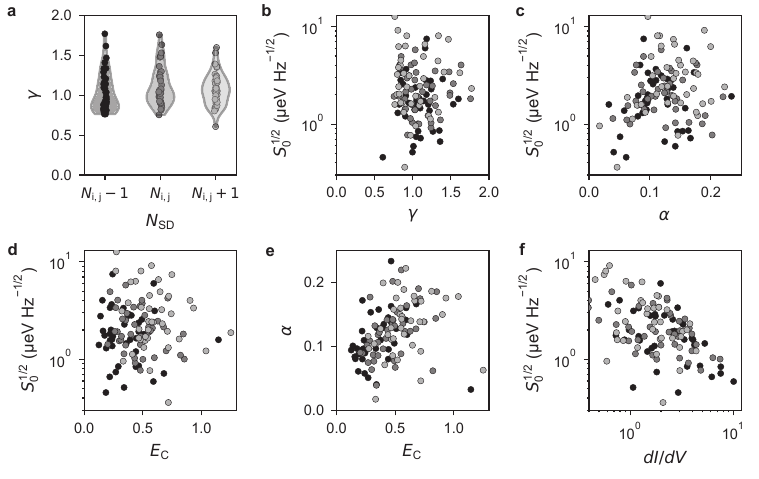}
    \caption{\textbf{a} Violin plots of the charge noise exponent $\gamma$ obtained from fitting the power spectral densities, showing that most values cluster around $\sim$ 1 \textbf{b} Scatter plot of $\gamma$ versus $S_\mathrm{0}^{1/2}$, revealing no apparent correlation between the two metrics, supporting a $1/f$ noise behaviour consistent with a log-uniform frequency distribution of TLFs. \textbf{c, d} Scatter plots of the lever arm ($\alpha$) and the charging energy ($E_C$) versus $S_\mathrm{0}^{1/2}$, respectively, both showing no significant correlation. \textbf{e} Scatter plot of $E_C$ versus $\alpha$, demonstrating a positive correlation, confirming that the quantum dot size influences the lever arm. Together with \textbf{c} and \textbf{d}, this suggests that dot size does not impact the magnitude of charge noise. \textbf{f} Scatter plot of the Coulomb peak slope versus $S_\mathrm{0}^{1/2}$, revealing a negative correlation, indicating that sharper peaks correspond to lower charge noise.}
    \label{figS8}
\end{figure*}
\newpage

\begin{figure*}[!ht]
    \centering
    \includegraphics[width=\linewidth]{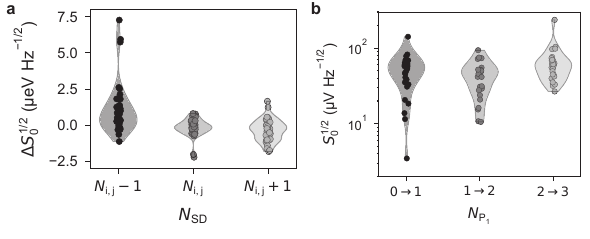}
    \caption{Violin plots for three consecutive hole occupancies in each tile reporting the distributions of the charge noise variations ($\Delta S_\mathrm{0}^{1/2}$) with respect to the average charge noise in each tile . With the exception of a few outliers, such as tile $(21,21)$, the median value of $\Delta S_\mathrm{0}^{1/2}$ fluctuates around zero, suggesting that in the multi-hole regime the specific charge occupation affects marginally the noise properties of the device. \textbf{b,} The violin plots of the charge noise measurements for 27 tiles across each charge occupation transition in $D_1$. The average charge noise does not depend significantly on hole filling, however, we observe a decrease in the standard deviation of the charge noise distribution with increasing hole occupation, indicating that higher hole occupations lead to more predictable average charge noise.}
    \label{figS9}
\end{figure*}
\newpage
\begin{figure*}[!ht]
    \centering
    \includegraphics[width=0.75\linewidth]{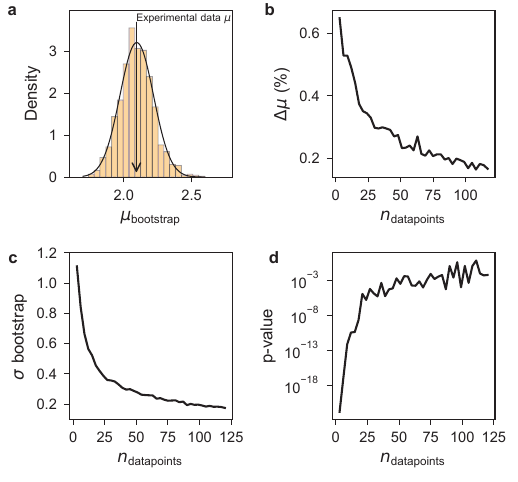}
    \caption{Histogram of bootstrapped means with the experimental mean from Fig.~4b indicated by a black arrow. The close agreement between the experimental and bootstrapped means (within 0.2\%) suggests that the experimental sample size (120 data points from 40 tiles) is sufficiently large to ensure an accurate estimation of the mean value. \textbf{b,} Difference between the bootstrapped and experimental means as a function of the experimental sample size. While the difference remains small throughout, it stabilizes at approximately 60 data points (20 tiles), indicating a sufficient sample size. \textbf{c,} Standard deviation of the bootstrapped means as a function of sample size, which similarly converges around 60 data points, reinforcing the reliability of the experimental sample size. \textbf{d,} p-value from a Shapiro-Wilk test as a function of sample size. This test assesses whether the dataset follows a normal distribution, which suggests that multiple independent random variables contribute additively to the measured quantity. High p-values ($>0.001$) indicate greater normality, which is achieved at approximately 20 tiles. Collectively, these results confirm that QARPET provides a sufficiently large dataset to accurately characterize charge noise.}
    \label{figS10}
\end{figure*}

\newpage
\begin{landscape}
    
\begin{figure*}[!ht]
    \centering
    \includegraphics[width=\linewidth]{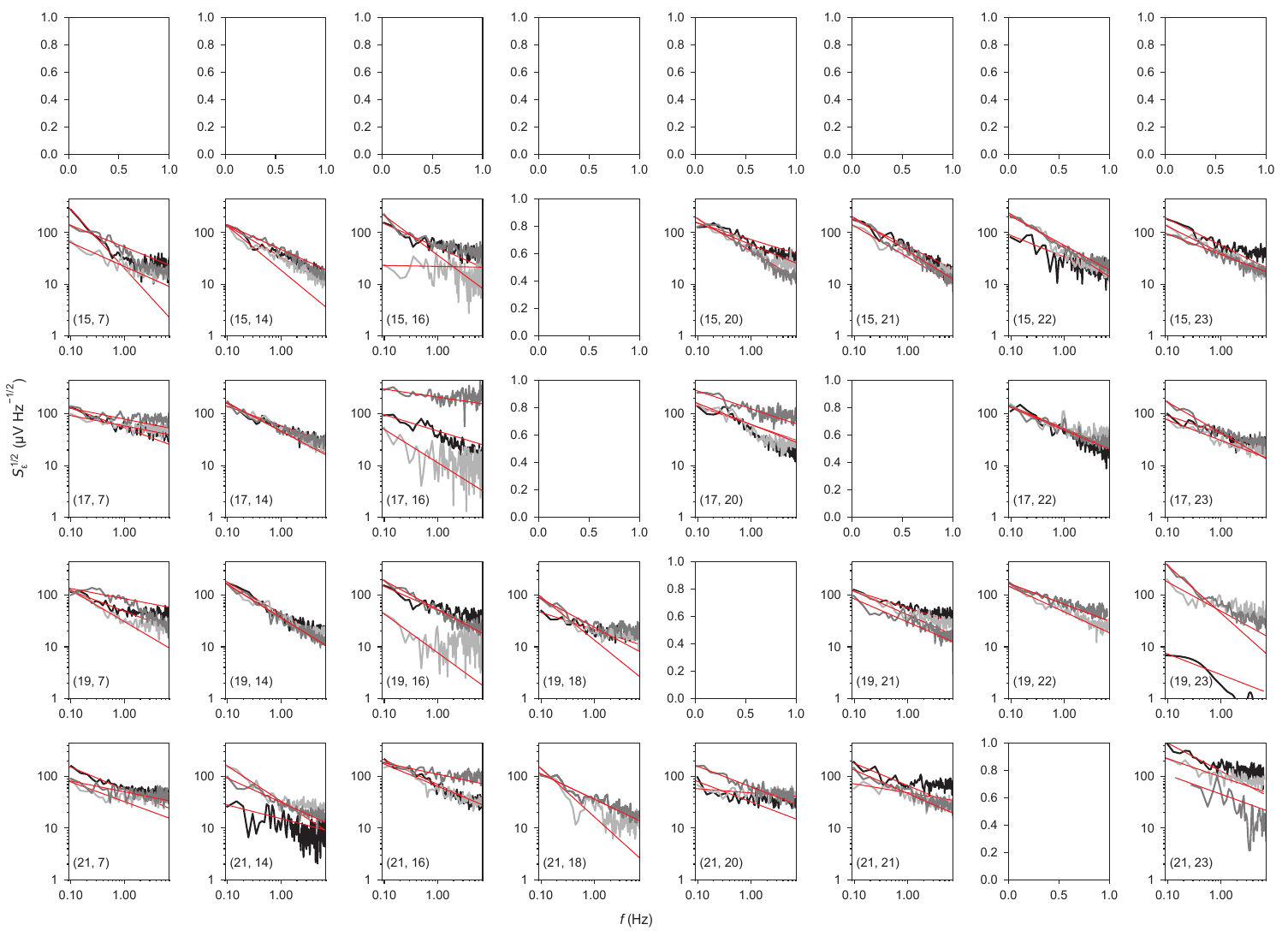}
    \caption{The voltage spectral densities $S_\epsilon^{1/2}$ from the three measured charge transitions of $D_1$ in each of the 27 investigated tiles. The spectral densities are individually fit to a $1/f^\gamma$ and both $S_0$ and $\gamma$ are extracted.}
    \label{figS11}
\end{figure*}

\end{landscape}

\newpage
\begin{figure*}[!ht]
    \centering
    \includegraphics[width=0.5\linewidth]{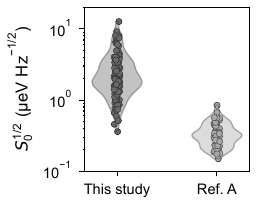}
    \caption{Violin plots of $S_\mathrm{0}^{1/2}$ at 1 Hz comparing the sensor noise of this study with data from Ref.~A (i.e Ref.~42 in the main text, \cite{stehouwer_exploiting_2025}) highlighting the significant differences in charge noise values between the two datasets.}
    \label{figS12}
\end{figure*}
\newpage
\begin{figure*}[!ht]
    \centering
    \includegraphics[width=0.5\linewidth]{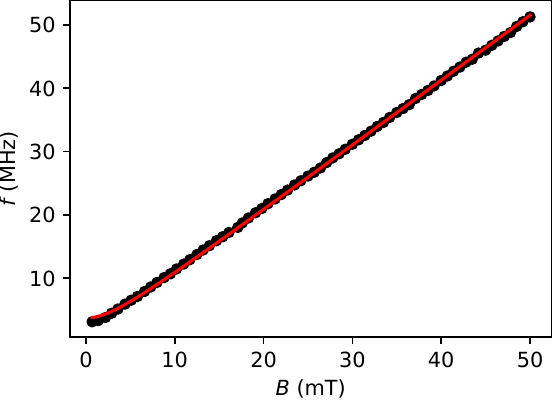}
    \caption{Frequency of the ST-oscillations in Fig.~5d at different magnetic fields. For each magnetic field we extract the frequency of the oscillations (fitting with a sine function) and the resulting points are fitted to $f = \frac{1}{h} \sqrt{J^2 + (\Delta g \mu_B B)^2}$ from which we extract the g-factor difference $\Delta g =0.0734 \pm 0.0001$ and the residual exchange at zero detuning $J(\varepsilon_{12}=0)= (3.736 \pm 0.112)$\SI{}{MHz}. Measurements are from QARPET device 2.}
    \label{fig:ST_fit}
\end{figure*}
\newpage

\section{Supplementary Note 1: Device Failure Modes}
\begin{itemize}

    \item\textbf{Electrode fractures:} Discontinuities in electrode lines result in localized failures affecting only the tiles in the specific row or column downstream of the interruption point. This failure mode has minimal impact on overall device functionality.

    \item\textbf{$P_s$, $B_s$ leakage:} Leakage of the sensor electrodes $P_s$ and $B_s$ produce failures constrained to the affected column or row, respectively. The spatial isolation of these failures allows for continued operation of unaffected device regions.

    \item\textbf{$B_1$, $P_1$, $B_2$, $P_2$ leakage:} If these lines are not shorted together during the bonding process, failures will be constrained to the affected column or row. If, to minimize the number of control lines, these lines are shorted together during the bonding process across the device (e.g., all $P_2$ shorted together), then a leakage affecting e.g. the $P_2$ line will make all $P_2$ non-operational across the entire device. However, to effectively mitigate this problem leakage can be measured at room temperature during the bonding process. If, after bonding a gate electrode, a small voltage is applied to it, there should be no current flowing (resistance should be typically greater than \SI{10}{\mega\ohm}). If the newly bonded electrode introduces leakage, then that bond can be removed. Note that bonding on the bonding pad should not create a path to ground via the substrate, and therefore care should be taken not to break the oxide under the bond by adjusting bonding force or adding an insulation layer under the bonding pads, for example SiN (see also Supplementary Fig.~1). One could also measure the resistance of each electrode vs. each other electrode, obtaining a matrix which indicates where the leakage occurs.
    
    \item \textbf{Localized defects:} impurities in the fabricated layers or defects in the material will only impact the affected tile.
\end{itemize}

\clearpage